\documentclass[10pt]{article}
\usepackage{amsmath,amssymb}
\usepackage{newtxtext,newtxmath,bm}
\usepackage{graphics,graphicx,epsfig}
\usepackage{natbib}
\usepackage[hmargin=1in,vmargin=1in]{geometry}
\numberwithin{equation}{section}

\begin{document}

%%%%%%%%%%%%%%%%%%%%%%%%%%%%%%%%%%%%%%%%%%%%%%%
%
\title{Coupled continuum modeling of size-segregation driven by shear-strain-rate gradients and flow in dense, bidisperse granular media}
\author{Daren Liu\footnotemark[2], Harkirat Singh\footnotemark[2], and David L. Henann\footnotemark[8]\\
School of Engineering, Brown University, Providence, RI 02912, USA}
\renewcommand*{\thefootnote}{\fnsymbol{footnote}}
\footnotetext[2]{These authors contributed equally to this work.}
\footnotetext[8]{Email address for correspondence: david\_henann@brown.edu}
\renewcommand*{\thefootnote}{\arabic{footnote}}
\date{}
\maketitle
\begin{abstract}
Dense granular systems that consist of particles of disparate sizes segregate based on size during flow, resulting in complex, coupled segregation and flow patterns. The ability to predict how granular mixtures segregate is important in the design of industrial processes and the understanding of geophysical phenomena. The two primary drivers of size-segregation are pressure gradients and shear-strain-rate gradients. In this work, we isolate size-segregation driven by shear-strain-rate gradients by studying two dense granular flow geometries with constant pressure fields: gravity-driven flow down a long vertical chute with rough parallel walls and annular shear flow with rough inner and outer walls. We perform discrete element method (DEM) simulations of dense flow of bidisperse granular systems in both flow geometries, while varying system parameters, such as the flow rate, flow configuration size, fraction of large/small grains, and grain-size ratio, and use DEM data to inform continuum constitutive equations for the relative flux of large and small particles. When the resulting continuum model for the dynamics of size-segregation is coupled with the nonlocal granular fluidity model--a nonlocal continuum model for dense granular flow rheology--we show that both flow fields and segregation dynamics may be simultaneously captured using this coupled, continuum system of equations. 
\end{abstract}
%
%%%%%%%%%%%%%%%%%%%%%%%%%%%%%%%%%%%%%%%%%%%%%%%

\section{Introduction}\label{sec_intro_seg}
Dense granular systems in nature and industry are often non-monodisperse--i.e., consisting of particles of disparate sizes. In non-monodisperse granular systems, the constituent grains segregate based on size during flow, forming complex patterns \citep[e.g.,][]{shinbrot2000,gray05,hill08,fan10,schlick2015,gray2018,umbanhowar2019modeling}. The ability to predict the dynamics of segregation is important across applications. For example, in geophysics, granular size segregation can manifest in landslides and debris flows \citep{johnson2012grain}, in which larger grains segregate to the top of the flow, potentially causing more damage, while in industry, size-segregation can be an undesirable effect when blending granular constituents of various sizes.

The current understanding is that there are two driving forces for size-segregation in dense granular flows. The first is pressure-gradients, which are typically induced by gravity. In pressure-gradient-driven size-segregation, small particles move more readily through the interstitial spaces that open and close during flow through a process referred to in the literature as ``kinetic sieving,'' leading to a system stratified along the direction of pressure gradients \citep{savage88,gray05,gray06,thornton12,fan14}. While pressure-gradient-driven segregation has been the focus of significant study, Hill and coworkers \citep{hill08,fan10,fan11b,hill2014} demonstrated that grains can also segregate in inhomogeneous flows along directions orthogonal to gravitationally-induced pressure gradients, driven instead by gradients in the shear-strain-rate. As an example, this mechanism has been observed in the split-bottom cell experiments of \citet{hill08}. In these experiments, not only do the larger particles segregate to the top of the cell, but they also segregate perpendicular to the direction of pressure gradients towards more rapidly shearing regions. Shear-strain-rate-gradient-driven segregation has received comparatively less attention in modeling efforts.

Due to the complexity of flow and segregation patterns, developing a general, predictive, continuum model for coupled size-segregation and flow in dense granular materials remains an open challenge. Although much progress has been made over recent decades \citep[e.g.,][]{savage88,gray05,gray06,fan11b,fan14,tunuguntla2017,gray2018,umbanhowar2019modeling}, the development of continuum models that are capable of simultaneously predicting the evolution of both segregation and flow fields, based solely on the geometry of the flow configuration, applied loads, and boundary/initial conditions is still in its infancy. Instead, most continuum models for size-segregation require some flow field quantity, such as the velocity or stress fluctuation field, to be measured first from experiments or DEM simulations and then used as model input. A crucial reason for the incompleteness of current models is the lack of a dense granular flow rheology theory that may be coupled to segregation models. A widely-used class of viscoplastic models for steady, dense granular flow is based on the $\mu(I)$ rheology \citep{midi2004,jop2005,dacruz2005,srivastava2021viscometric}. One recent work that couples rheology and segregation in dense granular flows is that of \citet{barker2021coupling}, which combines a regularized version of the $\mu(I)$  rheology \citep{barker2017partial} with a model for gravity-driven segregation. However, it has been well-documented in the literature \citep[e.g.,][]{kamrin2019non} that the $\mu(I)$ rheology, even in its regularized form, can break down in the presence of spatial flow inhomogeneity, which can be attributed to nonlocal effects not accounted for in the $\mu(I)$ rheology. 

To address this point, significant effort has gone into the development of size-dependent, nonlocal continuum constitutive theories for dense granular flow rheology, and coupling a nonlocal rheological model with a segregation model provides a route to robust, simultaneous prediction of flow and segregation fields. In this paper, we focus on the nonlocal granular fluidity (NGF) model of Kamrin and coworkers \citep{kamrin2012,henann2013,kamrin2019non}, which has been successfully applied to predicting dense flows of monodisperse grains in a wide variety of flow geometries. Then, the overarching aim of this paper is to formulate a predictive continuum theory for simultaneous flow and size-segregation in dense granular systems by integrating the NGF model with a phenomenological size-segregation model. This is a broad goal, and in this paper, we narrow our focus to several simpler, quasi-one-dimensional flow configurations. In most real-world flows, both the pressure-gradient-driven and shear-strain-rate-gradient-driven segregation mechanisms are present, making it difficult to disentangle them. Therefore, our plan for this paper is to isolate and examine the shear-strain-rate-gradient-driven mechanism. Specifically, we study the shear-strain-rate-gradient-driven segregation mechanism by considering flows of dense, bidisperse systems of both disks and spheres in flow geometries in which the pressure field is spatially uniform--namely, vertical chute flow and annular shear flow. Therefore shear-strain-rate-gradients are the sole drivers of segregation. In order to inform continuum model development, we perform discrete element method (DEM) simulations using the open source software LAMMPS \citep{lammps}, which function as ``numerical experiments.'' The coupled continuum model that we develop is then validated by comparing its predictions of the transient evolution of segregation and flow fields against additional DEM simulation results.

This paper is organized as follows. In Section \ref{sec_cont_seg}, we discuss the continuum model that we use to describe flow and size-segregation in bidisperse, dense granular materials. Specifically, Sections~\ref{subsec_mixt_seg} and \ref{subsec_eom} introduce the mixture theory framework used to describe dense, bidisperse granular mixtures, and in Section \ref{subsec_rheo_seg}, we briefly revisit the $\mu(I)$ rheology and the NGF model for monodisperse granular systems and discuss their extension to bidisperse systems. Then in Section \ref{subsec_segmod_seg}, we propose a model for shear-strain-rate-gradient-driven size-segregation. In Sections~\ref{sec_diff_seg} and  \ref{sec_shseg_seg}, we consider granular diffusion and shear-strain-rate-gradient-driven segregation, respectively, and independently determine the two dimensionless material parameters that appear in the size-segregation model for both disks and spheres. Then, in Section \ref{sec_shtrans_seg}, the proposed segregation model is coupled with the NGF model and applied to both vertical chute flow and annular shear flow to predict the transient evolution of the segregation dynamics, and the predicted continuum fields are compared against DEM measurements. In the end, our model demonstrates a level of fidelity in simultaneously predicting flow and segregation dynamics that has not been previously achieved.  We close with a discussion of the segregation model and some concluding remarks in Section \ref{sec_concl_seg}.

\section{Continuum framework}\label{sec_cont_seg}
In this section, we discuss the continuum framework used to describe dense, bidisperse granular systems and propose constitutive equations for rheology and size-segregation. Throughout, we utilize a mixture-theory-based approach, which is common in continuum modeling of dense, bidisperse mixtures \citep[e.g.,][]{gray05,gray06,fan11b,gray2018,umbanhowar2019modeling,barker2021coupling,bancroft2021drag,duan2021modelling}, and we use standard component notation, which supposes an underlying set of Cartesian basis vectors $\{{\bf e}_i|i=1,2,3\}$, and in which the components of vectors, ${\bf v}$, and tensors, $\boldsymbol{\sigma}$, are denoted by $v_i$ and $\sigma_{ij}$, respectively. The Einstein summation convention is employed, and the Kronecker delta, $\delta_{ij}$, is utilized to denote the components of the identity tensor.

\subsection{Bidisperse systems}\label{subsec_mixt_seg}
\begin{figure}[!t]
\begin{center}
\includegraphics{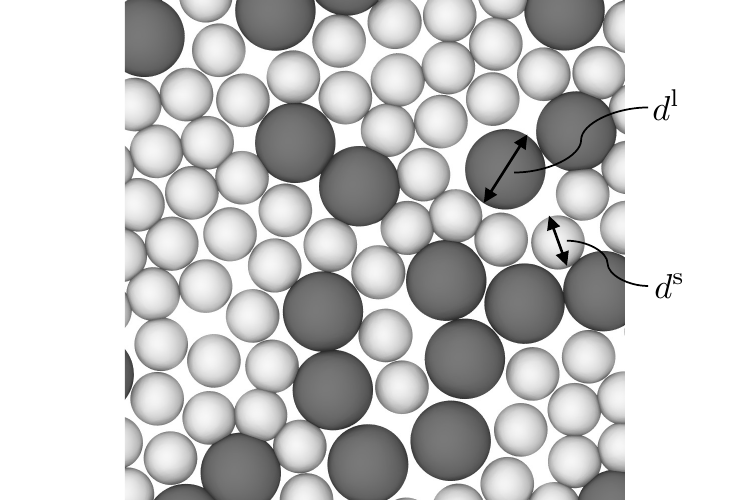}
\end{center}
\caption{A representative schematic of a dense, bidisperse granular system consisting of two-dimensional disks.}\label{fig1_SegS}
\end{figure}
We consider granular mixtures consisting of particles with two sizes--large grains with an average diameter of $d^{\rm l}$ and small grains with an average diameter of $d^{\rm s}$. We consider both two-dimensional systems of disks, as illustrated in Fig.~\ref{fig1_SegS}, and three-dimensional systems of spheres. To eliminate the effect of density-based segregation \citep[e.g.,][]{tripathi2013} and isolate size-based segregation, all particles are made of the same material with density $\rho_{\rm s}$, which represents the area-density for disks and the volume-density for spheres. Throughout, we utilize the notational convention in which we denote large-grain quantities using a superscript ${\rm l}$ and small-grain quantities using a superscript ${\rm s}$. The species-specific solid fractions--i.e., the areas occupied by each species per unit total area for disks and the volumes occupied by each species per unit total volume for spheres--are $\phi^{\rm l}$ and $\phi^{\rm s}$, respectively, and the total solid fraction is $\phi = \phi^{\rm l} + \phi^{\rm s}$. The concentration of each species then follows as $c^{\rm l} = \phi^{\rm l}/\phi$ and $c^{\rm s} = \phi^{\rm s}/\phi$, so that $c^{\rm l} + c^{\rm s} = 1$. The average mixture grain size is defined as the sizes of both species weighted by their concentrations, $\bar{d} = c^{\rm l}d^{\rm l} + c^{\rm s}d^{\rm s}$. We make the common idealization that the total area for dense systems of disks or total volume for dense systems of spheres does not change \citep{savage1998,gray05,gray06,fan11b}, and therefore $\phi$ is idealized as constant at each point in space and at each instant in time during the segregation process. We have verified in our DEM simulations that area (or volume) dilatation at flow initiation occurs over a much shorter time scale than the process of segregation, so that this idealization is reasonable. Throughout this study, we use $\phi=0.8$ for disks, and $\phi=0.6$ for spheres.

Regarding the kinematics of flow, each species has an associated partial velocity, $v_i^{\rm l}$ and $v_i^{\rm s}$, and the mixture velocity is given by $v_i = c^{\rm l}v_i^{\rm l} + c^{\rm s}v_i^{\rm s}$. The mixture strain rate tensor is then defined using the mixture velocity in the standard way: $D_{ij} = (1/2)(\partial v_i/\partial x_j + \partial v_j/\partial x_i)$, where $D_{kk}=0$ since we have assumed that the mixture area (or volume) does not change. The equivalent shear strain-rate is defined as $\dot\gamma = (2D_{ij}D_{ij})^{1/2}$.

Then, the relative area (or volume) flux for each grain type $\alpha={\rm l},{\rm s}$ is defined through the difference between its partial velocity and the mixture velocity as $w_i^{\alpha} = c^{\alpha} \left( v_i^{\alpha}-v_i \right),$ so that $w_i^{\rm l} + w_i^{\rm s}=0$. Conservation of mass for each species requires that $D{c}^{\alpha}/Dt + \partial w_i^{\alpha}/\partial x_i = 0$, where $D(\bullet)/Dt$ is the material time derivative. Due to the fact that $c^{\rm l} + c^{\rm s} = 1$, only one of $c^{\rm l}$ and $c^{\rm s}$ is independent. Therefore, we will utilize $c^{\rm l}$ as the field variable that describes the dynamics of size-segregation in the following discussion, and the evolution of $c^{\rm l}$ is governed by its conservation of mass equation:
\begin{equation}\label{mass_conserv_seg}
\dfrac{D{c}^{\rm l}}{Dt} + \dfrac{\partial w_i^{\rm l}}{\partial x_i} = 0.
\end{equation}

\subsection{Stress and the equations of motion}\label{subsec_eom}
We recognize that the symmetric Cauchy stress tensor $\sigma_{ij} = \sigma_{ji}$ represents the Cauchy stress of the mixture, rather than the partial stress of either species. Regarding stress-related quantities for the granular mixture, we define the pressure $P=-(1/3)\sigma_{kk}$, the stress deviator $\sigma_{ij}' = \sigma_{ij} + P\delta_{ij}$, the equivalent shear stress $\tau = (\sigma_{ij}'\sigma_{ij}'/2)^{1/2}$, and the stress ratio $\mu=\tau/P$. The Cauchy stress is then governed by the standard equations of motion
\begin{equation}\label{Chap_NGFmodel_eom}
\phi\rho_{\rm s}\frac{D v_i}{Dt}= \frac{\partial \sigma_{ij}}{\partial x_j} + b_i,
\end{equation}
where $\phi$ is the constant total solid fraction, and $b_i$ is the non-inertial body force per unit volume (typically gravitational). In order to close the system of equations, we require (1) rheological constitutive equations for the Cauchy stress $\sigma_{ij}$ and (2) a constitutive equation for the flux $w_i^{\rm l}$, each of which are discussed in the following subsections.

\subsection{Rheological constitutive equations for bidisperse mixtures}\label{subsec_rheo_seg}
In this section, we discuss the rheology of dense, bidisperse granular mixtures.  Our strategy for formulating rheological constitutive equations for bidisperse mixtures is to relate mixture-related quantities, such as the Cauchy stress $\sigma_{ij}$ and the strain-rate tensor $D_{ij}$, instead of specifying constitutive equations for species-specific partial stresses and then combining them to obtain the mixture stress. 

The starting point of this discussion is the local inertial, or $\mu(I)$, rheology \citep{midi2004,jop2005,dacruz2005}, which follows from dimensional arguments. For a dense, monodisperse system of dry, stiff grains with mean grain diameter $d$ subjected to homogeneous shearing, the local inertial rheology asserts that the stress ratio $\mu$ is given through the equivalent shear strain-rate $\dot\gamma$ and the pressure $P$ through the  dimensionless relationship $\mu=\mu_{\rm loc}(I)$, where $I=\dot\gamma\sqrt{d^2\rho_{\rm s}/P}$ is the inertial number, representing the ratio of the microscopic time scale associated with particle motion $\sqrt{d^2\rho_{\rm s}/P}$ to the macroscopic time scale of applied deformation $1/\dot\gamma$. As shown by \citet{rognon07} and \citet{tripathi11}, the inertial rheology function $\mu_{\rm loc}(I)$ may be straightforwardly generalized from monodisperse to bidisperse systems by defining the inertial number for a bidisperse system as $I = \dot\gamma\sqrt{\bar{d}^2\rho_{\rm s}/P}$, where the average mixture grain size for a bidisperse system $\bar{d}$ has been used in place of $d$ for a monodisperse system. Then, the same local rheology function $\mu_{\rm loc}(I)$ utilized for the monodisperse system may be used for bidisperse systems without any changes to the parameters appearing in the fitting function. This approach neglects potential effects of new dimensionless quantities that arise in a bidisperse granular system, such as the grain size ratio $d^{\rm l}/d^{\rm s}$, but has been shown to capture DEM data well \citep{rognon07,tripathi11}.

\begin{figure}[!t]
\begin{center}
\includegraphics{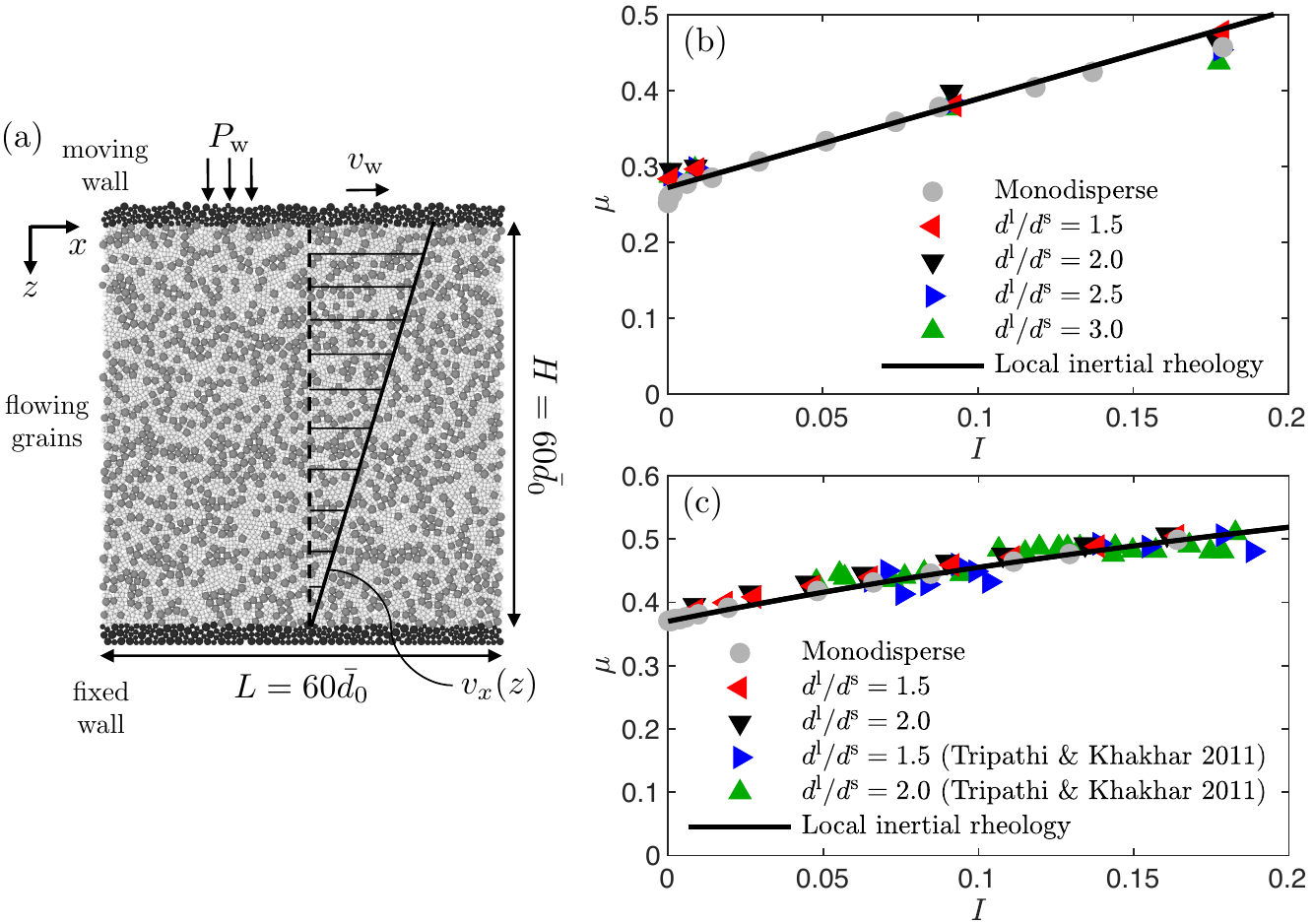}
\end{center}
\caption{(a) Configuration for two-dimensional DEM simulations of bidisperse simple shear flow. Upper and lower layers of black grains denote rough walls. Dark gray grains indicate large flowing grains, and light gray grains indicate small flowing grains. A $10\%$ polydispersity is utilized for each species to prevent crystallization. (b) The local inertial rheology ($\mu$ versus $I = \dot\gamma\sqrt{\bar{d}^2\rho_{\rm s}/P}$) for monodisperse as well as bidisperse mixtures of disks for grain-size ratios of $d^{\rm l}/d^{\rm s} = 1.5$, 2.0, 2.5, and 3.0 and $c^{\rm l}=0.5$. The solid black line represents the best fit to the monodisperse DEM data using \eqref{local_seg} with $\mu_{\rm s}=0.272$ and $b=1.168$. (c) The local inertial rheology for monodisperse and bidisperse mixtures of spheres for grain size ratios of $d^{\rm l}/d^{\rm s} = 1.5$ and 2.0 and $c^{\rm l} = 0.5$ along with the DEM data of \citet{tripathi11}. The solid black curve represents the best fit to the monodisperse DEM data using \eqref{local_seg2} with $\mu_{\rm s}=0.37$, $\mu_2 = 0.95$, and $I_0=0.58$.}\label{fig2_SegS}
\end{figure}

To demonstrate this point, consider DEM simulations of homogeneous, simple shearing of a dense, bidisperse system of disks, illustrated in Fig.~\ref{fig2_SegS}(a) for the case of $d^{\rm l}/d^{\rm s}=1.5$ and a system-wide large-grain concentration of $c^{\rm l}=0.5$. Details of the simulated granular systems, including grain interaction properties, for both two-dimensional disks and three-dimensional spheres are given in Appendix~\ref{app_granularsystem}. The large particles are dark gray, and the small particles are light gray. With the system-wide mean grain size denoted by $\bar{d}_0 = c^{\rm l}d^{\rm l} + (1-c^{\rm l})d^{\rm s}$, the rectangular domain has a length of $L=60\bar{d}_0$ in the $x$-direction and a height of $H=60\bar{d}_0$ in the $z$-direction, which is filled with $\sim 5000$ flowing grains. Shearing is driven through the relative motion of two parallel, rough walls, which each consist of a thin layer of touching glued grains, which are denoted as black in Fig.~\ref{fig2_SegS}(a). The bottom wall is fixed, and the top wall moves with a velocity $v_{\rm w}$ along the $x$-direction. Following previous works in the literature \citep{dacruz2005,koval2009,kamrin2012,zhang2017,liu2018,kim2020}, the $z$-position of the top wall is not fixed but continuously adjusted using a feedback scheme so that the normal stress applied by the top wall is maintained at a target value of $\sigma_{zz}(z=0)=-P_{\rm w}$. Periodic boundary conditions are applied along the $x$-direction. For homogeneous simple shearing, no segregation will occur since the flow is homogeneous and no pressure or strain-rate gradients are present. We utilize the DEM procedures described in detail in \citet{liu2018} in order to extract the relationship between $\mu$ and $I$ for bidisperse mixtures with grain size ratios of $d^{\rm l}/d^{\rm s} = 1.5, 2.0, 2.5$, and 3.0 and $c^{\rm l}=0.5$. The simulated relationships are plotted in Fig.~\ref{fig2_SegS}(b) using triangular symbols of different colors, along with the monodisperse data from \citet{liu2018} plotted as gray circles. The relationship between $\mu$ and $I$ for dense systems of disks is observed to be approximately independent of $d^{\rm l}/d^{\rm s}$. As for the monodisperse case, the DEM data for bidisperse mixtures of disks may be fit by a linear, Bingham-like functional form:  
\begin{equation}\label{local_seg}
\mu_{\rm loc}(I) = \mu_{\rm s} + bI,
\end{equation}
as shown by the solid line in Fig.~\ref{fig2_SegS}(b), where $\mu_{\rm s}=0.272$ and $b=1.168$ are the dimensionless material parameters for monodisperse disks \citep{liu2018}. 

Similarly, we consider DEM simulations of homogeneous, simple shearing of dense, bidisperse systems of spheres. The simulation domain consists of a rectangular box of length $L=20\bar{d}_0$ in the $x$-direction (i.e., the shearing direction), width $W=10\bar{d}_0$ in the $y$-direction (i.e., the direction perpendicular to the plane of shearing), and height $H=40\bar{d}_0$ in the $z$-direction. The domain is filled with $ \sim$10000 flowing grains, and periodic boundary conditions are applied along both the $x$- and $y$-directions. The simulation domain is bounded along the $z$-direction by two parallel, rough walls, consisted of touching glued grains, and as for the case of disks, shearing along the $x$-direction and normal stress along the $z$-direction are applied by the walls. We perform DEM simulations of steady simple shearing for size ratios of $d^{\rm l}/d^{\rm s} = 1.5$ and $2.0$ for a system-wide large-grain concentration of $c^{\rm l}=0.5$ as well as for the monodisperse case over a range of top wall velocities. The $\mu$ versus $I$ relationship extracted from DEM simulations for these cases along with data from the prior DEM study of \citet{tripathi11} collapse quite well as shown in Fig.~\ref{fig2_SegS}(c), showing minimal dependence on $d^{\rm l}/d^{\rm s}$. This relationship for dense systems of spheres may be fitted using a  nonlinear functional form of \citet{jop2005} for $\mu_{\rm loc}(I)$:
\begin{equation}\label{local_seg2}
 \mu_{\rm loc}(I) = \mu_{\rm s} + \frac{\mu_2 - \mu_{\rm s}}{I_0/I + 1},
\end{equation}
as shown by the solid curve in Fig.~\ref{fig2_SegS}(c), where $\{\mu_{\rm s} = 0.37,\, \mu_2 = 0.95, \, I_0 = 0.58\}$ are the dimensionless parameters for frictional spheres. (We note that these parameters are nearly the same as those determined by \citet{zhang2017} for monodisperse frictional spheres.) In this way, one may capture the rheology of bidisperse mixtures of both disks and spheres in homogeneous simple shearing without introducing additional fitting functions or adjustable parameters beyond those used for the monodisperse case.

Despite the successes of the local inertial rheology in capturing steady, homogeneous shear flow, it has been well established in the literature that a local rheological modeling approach cannot be applied to a broad set of inhomogeneous flows, such as annular shear flow \citep{tang2018}, split-bottom flow \citep{fenistein2003}, and gravity-driven heap flow \citep{komatsu2001}. Therefore, to consider inhomogeneous flows of bidisperse granular systems, it is necessary to generalize a nonlocal rheological modeling approach to the case of dense, bidisperse mixtures. In the present work, we focus attention on the nonlocal granular fluidity (NGF) model of \citet{kamrin2012}, which has been shown to robustly capture a variety of inhomogeneous, steady flows of monodisperse granular systems \citep{kamrin2019non}. As is standard in the NGF model, we introduce the granular fluidity $g$, which is a positive, scalar field quantity, and recognize that $g$ represents the fluidity of the mixture. (See \citet{zhang2017} and \citet{kim2020} for further discussion of the kinematic description of the granular fluidity field for monodisperse granular systems.) Then, we utilize the steady-state form of the NGF model, which relates the stress state, the strain-rate, and the granular fluidity through two constitutive equations: (1) the flow rule and (2) the nonlocal rheology. 

First, invoking the common idealization that the Cauchy stress deviator and the strain-rate tensor are co-directional \citep{rycroft2009}, the flow rule relates the Cauchy stress tensor $\sigma_{ij}$, the strain-rate tensor $D_{ij}$, and the granular fluidity through
\begin{equation}\label{flowrule3D}
\sigma_{ij} = -P\delta_{ij} + 2\dfrac{P}{g}D_{ij}.
\end{equation}
Taking the magnitude of the deviatoric part of \eqref{flowrule3D} and rearranging leads to the following scalar form of the flow rule:
\begin{equation}\label{flowrule1D_seg}
\dot\gamma = g\mu.
\end{equation}

Second, the granular fluidity of the bidisperse mixture is governed by the following differential relation:
\begin{equation}\label{nonlocal_seg}
g = g_{\rm loc}(\mu,P) + \xi^2(\mu) \frac{\partial^2 g}{\partial x_i \partial x_i},
\end{equation}
where $g_{\rm loc}(\mu,P)$ is the local fluidity function and $\xi(\mu)$ is the stress-dependent cooperativity length. The local fluidity function gives the granular fluidity during steady, homogeneous shear flow at a given state of stress and is related to the local inertial rheology function $\mu_{\rm loc}(I)$. Denote the inverted form of the local inertial rheology function $\mu_{\rm loc}(I)$ as 
\begin{equation}\label{Iloc}
 I_{\rm loc}(\mu) = \left\{\begin{array}{cl}
\mu^{-1}_{\rm loc}(\mu) & \text{if $\mu>\mu_{\rm s}$,}\\[4pt]
0 & \text{if $\mu\le \mu_{\rm s}$,}\end{array}\right.
\end{equation}
which is a function of the stress ratio $\mu$. Then, consistent with the new definition of the inertial number involving $\bar{d}$ for a bidisperse mixture, the local fluidity function is $g_{\rm loc}(\mu,P) = \sqrt{P/\bar{d}^2\rho_{\rm s}}\, I_{\rm loc}(\mu)/\mu$. For the case of bidisperse disks, using \eqref{local_seg}, the local fluidity function is 
\begin{equation}\label{localg_bi_seg}
g_{\rm loc}(\mu,P) 
=  \left\{\begin{array}{cl}\sqrt{\dfrac{P}{\rho_{\rm s}\bar{d}^2}} \, \dfrac{(\mu-\mu_{\rm s})}{b\mu}  &\text{if $\mu>\mu_{\rm s}$,}\\[8pt]
0 &\text{if $\mu\le\mu_{\rm s}$,}\end{array}\right.
\end{equation}
with $\{\mu_{\rm s}=0.272, b=1.168\}$, and for the case of bidisperse spheres, using \eqref{local_seg2}, the local fluidity function is
\begin{eqnarray}\label{localg_bi_seg_spheres}
g_{\rm loc}(\mu,P) = \left\{
\begin{array}{cl}
 I_0\sqrt{\dfrac{P}{\rho_{\rm s}\bar{d}^2}}\,\dfrac{(\mu-\mu_{\rm s})}{\mu(\mu_2-\mu)} & \text{if $\mu>\mu_{\rm s}$,}\\[8pt]
 0 & \text{if $\mu\le\mu_{\rm s}$,}\\
\end{array}
\right.
\end{eqnarray}
with $\{\mu_{\rm s} = 0.37,\, \mu_2 = 0.95, \, I_0 = 0.58\}$. No additional adjustable parameters beyond those used to describe the local inertial rheology for the monodisperse case are introduced in the local fluidity function.

As discussed in several of our previous works \citep{henann2014,kamrin2015,liu2017}, the functional form for the cooperativity length $\xi(\mu)$ is also connected to the choice of the $\mu_{\rm loc}(I)$ function. Without going into details here, the functional forms for the cooperativity length corresponding to \eqref{local_seg} and \eqref{local_seg2} are
\begin{equation}\label{cooperativity_bi_seg}
\xi(\mu) = \frac{A\bar{d}}{\sqrt{|\mu - \mu_{\rm s}|}}\quad\text{and}\quad \xi(\mu) = A\bar{d}\sqrt{\dfrac{(\mu_2-\mu)}{(\mu_2-\mu_{\rm s})|\mu-\mu_{\rm s}|}}, 
\end{equation}
respectively. In the monodisperse case, the cooperativity length is directly proportional to the grain size $d$, and in \eqref{cooperativity_bi_seg} for the bidisperse case, it is taken to be proportional to $\bar{d}$. This is analogous to the process undertaken above for generalizing the local inertial rheology, in which $d$ for monodisperse grains is replaced by $\bar{d}$ for bidisperse grains. Moreover, in \eqref{cooperativity_bi_seg}, the parameter $A$ is a dimensionless material constant, referred to as the non-local amplitude, which quantifies the spatial extent of cooperative effects. Again, following an approach analogous to the generalization of the local inertial rheology, we utilize values of $A$ previously determined for monodisperse frictional disks and spheres--namely, $A=0.9$ as determined by \citet{liu2018} for monodisperse disks and $A=0.43$ as determined by \citet{zhang2017} for monodisperse spheres. These choices of $A$ for bisdisperse mixtures will be tested in later sections by comparing flow fields predicted by the NGF model to measured  flow fields in DEM simulations of bidisperse, inhomogeneous flows.

\subsection{Segregation model}\label{subsec_segmod_seg}
The segregation model consists of the constitutive equation for the large-grain flux $w_i^{\rm l}$. In the present work, we focus on dense flows in the absence of pressure gradients, and we take the large-grain flux $w_i^{\rm l}$ to be comprised of two contributions: (1) a diffusion flux $w_i^{\rm diff}$ and (2) a shear-strain-rate-gradient-driven segregation flux $w_i^{\rm seg}$, so that
\begin{equation}\label{flux_decomp_seg}
w_i^{\rm l} = w_i^{\rm diff} + w_i^{\rm seg}.
\end{equation}

First, the diffusion flux acts counter to segregation to mix the species and is taken to be given in the standard form, in which the diffusion flux is driven by concentration gradients: $w_i^{\rm diff} = -D \left( \partial c^{\rm l}/\partial x_i \right)$, where $D$ is the binary diffusion coefficient \citep{utter2004,artoni2021self,bancroft2021drag}. Based on dimensional arguments, we expect that 
\begin{equation}\label{diff_coeff_seg}
D = C_{\rm diff} \bar{d}^2 \dot{\gamma},
\end{equation}
where $C_{\rm diff}$ is a dimensionless material parameter which remains to be calibrated \citep{fan14,tripathi2013}. Therefore, we have that the diffusion flux is 
\begin{equation}\label{diff_flux_seg}
w_i^{\rm diff} = -C_{\rm diff} \bar{d}^2 \dot{\gamma} \dfrac{\partial c^{\rm l}}{\partial x_i}.
\end{equation}

Second, regarding segregation, a major question is what field quantity drives the segregation flux in the absence of pressure gradients. Gradients of a number of kinematic quantities are possible--e.g., strain-rate, velocity fluctuations, or fluidity. For perspective, we note that recent works  \citep{fan11b,hill2014,tunuguntla2016,tunuguntla2017} have shown that gradients in kinetic stress, which are defined through the velocity fluctuations, correlate well with segregation flux. In the present work, we adopt the simplest approach and hypothesize that the segregation flux is driven by gradients in the shear strain-rate $\dot\gamma$ and take the segregation flux to be given in the following phenomenological form:
\begin{equation}\label{segS_flux_seg}
w_i^{\rm seg} = C_{\rm seg}^{\rm S} \bar{d}^2 c^{\rm l}(1-c^{\rm l}) \dfrac{\partial \dot{\gamma}}{\partial x_i}.
\end{equation}
The factor $c^{\rm l}(1-c^{\rm l})$ ensures that segregation ceases when the bidisperse mixture becomes either all large $(c^{\rm l} = 1)$ or all small $(c^{\rm l} = 0)$ grains, and the factor $\bar{d}^2$ is present for dimensional consistency. The quantity $C^{\rm S}_{\rm seg}$ is a dimensionless material property. While it is possible for $C^{\rm S}_{\rm seg}$ to depend on the size ratio $d^{\rm l}/d^{\rm s}$, we will demonstrate that this effect is negligible in the DEM simulations of Section~\ref{sec_shseg_seg} and therefore treat $C^{\rm S}_{\rm seg}$ as a constant, dimensionless material parameter, which will be determined by fitting to DEM simulation results for disks and spheres, respectively.

Combining \eqref{diff_flux_seg}, \eqref{segS_flux_seg}, and \eqref{flux_decomp_seg} with conservation of mass \eqref{mass_conserv_seg}, we obtain the following differential relation governing the dynamics of $c^{\rm l}$:
\begin{equation}\label{segS_model_seg}
\dfrac{D{c}^{\rm l}}{Dt} + \dfrac{\partial}{\partial x_i} \left( -C_{\rm diff} \bar{d}^2 \dot{\gamma} \dfrac{\partial c^{\rm l}}{\partial x_i} + C^{\rm S}_{\rm seg} \bar{d}^2 c^{\rm l}(1-c^{\rm l}) \dfrac{\partial \dot{\gamma}}{\partial x_i} \right) = 0 ,
\end{equation}
where $\{ C_{\rm diff}, C^{\rm S}_{\rm seg} \}$ represent  two constant dimensionless material parameters that remain to be determined.

We close this section by noting that the incompressibility constraint, the equations of motion \eqref{Chap_NGFmodel_eom}, the nonlocal rheology \eqref{nonlocal_seg}, and the segregation dynamics equation \eqref{segS_model_seg} represent a closed system of equations for the velocity field $v_i$, the pressure field $P$, the fluidity field $g$, and the large-grain concentration field $c^{\rm l}$, which may be used to simultaneously predict flow fields and segregation dynamics in the absence of pressure gradients. 

\section{Diffusion flux}\label{sec_diff_seg}
In this section, we determine  values of $C_{\rm diff}$ for dense, bidisperse systems of frictional disks and spheres. Consider homogeneous simple shear flow of such a bidisperse mixture, as shown in Fig.~\ref{fig2_SegS}(a) for disks. Again, no segregation occurs in this setting, since neither of the segregation driving forces (pressure gradients or shear-strain-rate gradients) are present \citep{tripathi11}. During steady, simple shearing, the motion of individual grains in the direction transverse to flow (the $z$-direction in Fig.~\ref{fig2_SegS}(a)) approximates a random walk for both two-dimensional systems of disks and three-dimensional systems of spheres. Therefore, by measuring the mean square displacement (MSD) of the system of $N$ particles as a function of time, we may determine the binary diffusion coefficient $D$  \citep{natarajan1995,dufty2002,utter2004,fan2015,kharel2017,bancroft2021drag} through 
\begin{equation}\label{msd_seg}
\text{MSD}(t)=\dfrac{1}{N} \sum_{n=1}^N(z_n(t)-z_n(0))^2 = 2Dt,
\end{equation}
where $z_n(t)$ is the $z$-coordinate of the $n$th grain at time $t$. We simulate homogeneous, steady simple shear flows of disks for grain-size ratios of $d^{\rm l}/d^{\rm s} = 1.5$, 2.0, 2.5, 3.0, and 4.0 and at various shearing rates. We also simulate homogeneous, steady simple shear flows of spheres for the monodisperse case as well as for grain-size ratios of $d^{\rm l}/d^{\rm s} = 1.5$, 2.0, and 2.5 over a range of shearing rates. To avoid wall effects in the calculation of the MSD \eqref{msd_seg}, grains that are initially within $15\bar{d}_0$ of either the top or bottom wall in Fig.~\ref{fig2_SegS}(a) are excluded from the system of particles used to calculate the MSD for disks, leaving a set of $N\approx 2400$ grains. For spheres, particles initially within $5\bar{d}_0$ of the top and bottom walls are excluded, so that a set of $N\approx 9000$ grains are used to calculate the MSD. Both large and small grains are included in the calculation of the MSD for the mixture. After a sufficiently long time, the calculated MSD is linear in time in all cases for both disks and spheres, allowing one to extract the diffusion coefficient $D$ for each case.

\begin{figure}[!t]
\begin{center}
\includegraphics{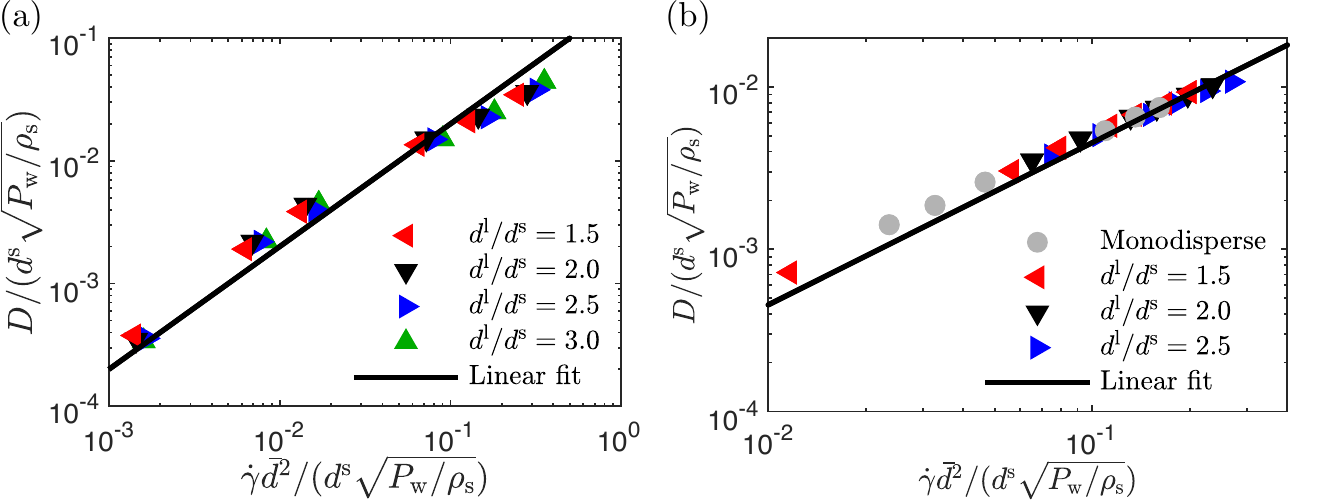}
\end{center}
\caption{The binary diffusion coefficient $D$, calculated using the mean square displacement \eqref{msd_seg},  versus $\dot{\gamma}\bar{d}^2$ in homogeneous, steady simple shear DEM simulations. (a) Simple shearing of bidisperse mixtures of disks for grain-size ratios of $d^{\rm l}/d^{\rm s} = 1.5, 2.0, 2.5, 3.0$, and 4.0. Both axes are normalized by $d^{\rm s} \sqrt{P_{\rm w}/\rho_{\rm s}}$. Each symbol represents $D$ calculated from one DEM simulation of a specified size ratio at one shearing rate. The solid line represents the best fit of a linear relation with $C_{\rm diff}=0.20$. (b) Simple shearing of bidisperse mixtures of spheres for the monodisperse case and for grain-size ratios of $d^{\rm l}/d^{\rm s} = 1.5, 2.0$, and 2.5. The solid line represents the best fit of a linear relation with $C_{\rm diff}=0.045$.}\label{fig3_SegS}
\end{figure}

The diffusion coefficient $D$ is plotted  against $\dot{\gamma}\bar{d}^2$ (with both quantities normalized by $d^{\rm s} \sqrt{P_{\rm w}/\rho_{\rm s}}$) for disks in Fig.~\ref{fig3_SegS}(a) and for spheres in Fig.~\ref{fig3_SegS}(b). The DEM data for the binary diffusion coefficient collapses to a nearly linear relation with $D\sim \dot{\gamma}\bar{d}^2$ across the range of size ratios and shearing rates considered. A best fit of the slopes of the linear relations--the solid black lines in Figs.~\ref{fig3_SegS}(a) and (b)--yields:
\begin{equation}\label{fit_Cdiff_seg}
C_{\rm diff}=\dfrac{D}{\dot{\gamma}\bar{d}^2}=0.20\text{ for disks and }C_{\rm diff}=\dfrac{D}{\dot{\gamma}\bar{d}^2}=0.045\text{ for spheres.}
\end{equation}
We note that our results are consistent with previous results in the literature. For example, the recent work of \citet{bancroft2021drag} found a value of $C_{\rm diff}\approx 0.05$ with a weak dependence on the inertial number for dense spheres. In order to further assess the fitted value of $C_{\rm diff}$ in a diffusion-dominated setting, we have performed a consistency test for disks by considering simple shearing of an initially fully-segregated cell, which is described in Appendix~\ref{app_diff_seg}. In this case, diffusion drives remixing of the two species. Using the fitted value of $C_{\rm diff}$ for disks in \eqref{fit_Cdiff_seg}, we are able to quantitatively capture the diffusive remixing process, which provides confidence in our fitted value of $C_{\rm diff}$.

\section{Shear-strain-rate-gradient-driven segregation flux}\label{sec_shseg_seg}
Having independently determined the material parameter $C_{\rm diff}$ for both frictional disks and spheres, we next turn to testing the constitutive equation for the shear-strain-rate-gradient-driven segregation flux \eqref{segS_flux_seg} and determining the material parameter $C^{\rm S}_{\rm seg}$ by studying two representative flow configurations in the absence of pressure gradients: (1) vertical chute flow and (2) annular shear flow.

\subsection{Vertical chute flow}\label{subsec_segVCF_seg}
\begin{figure}[!t]
\begin{center}
\includegraphics{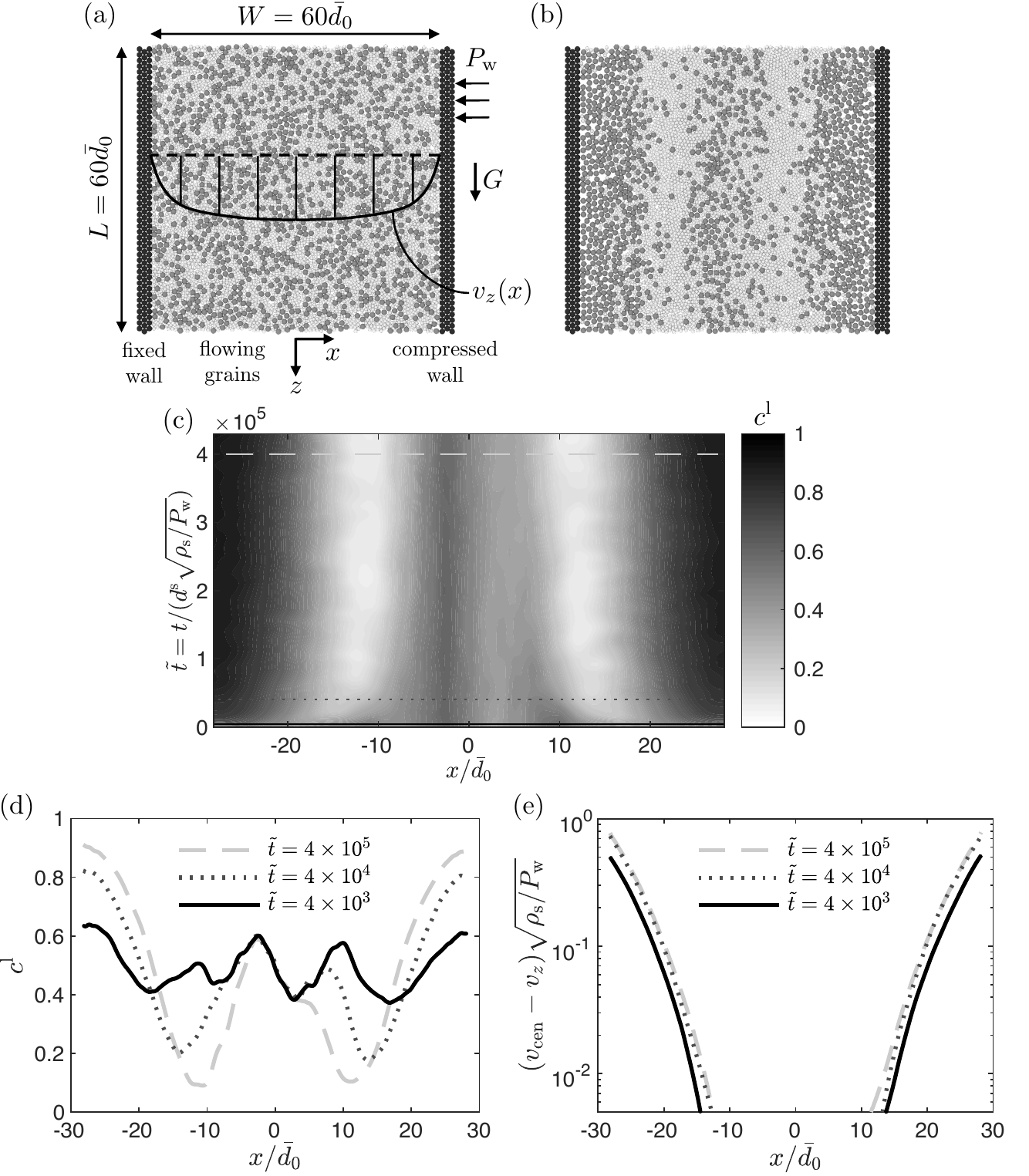}
\end{center}
\caption{(a) Initial well-mixed configuration for two-dimensional DEM simulation of bidisperse vertical chute flow with $4327$ flowing grains. The chute width is $W=60\bar{d}_0$. As in Fig.~\ref{fig2_SegS}, black grains on both sides represent rough walls (only large particles are used as wall grains here). (b) Segregated configuration after flowing for a total simulation time of $\tilde{t} = t/\left( d^{\rm s}\sqrt{\rho_{\rm s}/P_{\rm w}} \right)=4.3 \times 10^5$. (c) Spatiotemporal evolution of the large grain concentration field. Spatial profiles of (d) the concentration field $c^{\rm l}$ and (e) the normalized velocity field $(v_{\rm cen}-v_z)\sqrt{\rho_{\rm s}/P_{\rm w}}$ at three times ($\tilde{t}  =4 \times 10^3$, $4 \times 10^4$ and $4 \times 10^5$) as indicated by the horizontal lines in (c).}\label{fig4_SegS}
\end{figure}

Consider a dense, bidisperse granular mixture flowing down a long vertical chute with parallel, rough walls separated by a distance $W$ under the action of gravity $G$. This flow geometry has been utilized extensively in the literature to study dense flows of monodisperse, frictional disks \citep{kamrin2012,liu2018} and spheres \citep{zhang2017,kim2020} as well as flows of bidisperse, frictional spheres \citep{fan11a,fan11b}. Beginning with the case of bidisperse disks, the DEM setup is shown in Fig.~\ref{fig4_SegS}(a) for $W=60\bar{d}_0$, where $\bar{d}_0$ is the system-wide average grain size. In all cases for disks, we take the chute length to be $L=60\bar{d}_0$ and apply periodic boundary conditions along the $z$-direction. The parallel, rough walls consist of touching glued large grains, denoted as black in Fig.~\ref{fig4_SegS}(a). The left vertical wall is fixed, and the right wall is fixed in the $z$-direction but can move slightly in the $x$-direction so as to maintain a constant compressive normal stress $P_{\rm w}$ on the granular material, utilizing the same wall-position control method described in \citet{liu2018}. We have verified that the chute length $L$ is sufficiently large, so that it does not affect the resulting flow and segregation fields and all fields are invariant along the $z$-direction. In the resulting flow fields, the only non-zero component of the velocity is $v_z$, which only depends on the cross-channel coordinate $x$. A typical steady velocity field is qualitatively sketched in Fig.~\ref{fig4_SegS}(a), illustrating that the shear-strain-rate is greatest at the walls $(x = \pm W/2)$.

In all of our DEM simulations of vertical chute flow of bidisperse disks, we observe that the stress field quickly becomes independent of time, so that macroscopic inertia (i.e., the left-hand-side of \eqref{Chap_NGFmodel_eom}) may be neglected. Moreover, we observe that the normal stresses are approximately equal, i.e., $\sigma_{zz}\approx\sigma_{xx}$. Therefore, due to the force balance along the $z$-direction, the equivalent shear stress field is $\tau(x) = |\sigma_{xz}(x)| = |\sigma_{zx}(x)| = \phi\rho_{\rm s}G|x|$, where $x$ is measured from the centerline of the chute, and due to the force balance along the $x$-direction, the pressure field is $P(x) = -\sigma_{xx}(x) = P_{\rm w}$. The stress ratio field then follows as  
\begin{equation}\label{VCF_stress_seg}
\mu(x) = \mu_{\rm w}\left(\dfrac{|x|}{W/2}\right),
\end{equation}
where $\mu_{\rm w}=\phi\rho_{\rm s}GW/2P_{\rm w}$ is the maximum value of $\mu$, occurring at the walls $(x = \pm W/2)$. We note that while flow is driven by gravity, the pressure field is constant throughout the chute, and no pressure gradients are present. Therefore, segregation occurs only due to shear-strain-rate-gradients, enabling us to consider this effect in isolation.

Apart from the grain interaction properties that are held constant throughout this work (Appendix~\ref{app_granularsystem}), there are four important dimensionless parameters that fully describe each case of vertical chute flow of dense, bidisperse granular mixtures: (1) $W/\bar{d}_0$, the dimensionless chute width; (2) $\mu_{\rm w}$, the maximum stress ratio, which occurs at the walls and controls the total flow rate; (3) $c^{\rm l}_0(x)$, the initial large-grain concentration, which is not necessarily constant but can be a spatially-varying field; and (4) $d^{\rm l}/d^{\rm s}$, the bidisperse grain-size ratio. This list of system parameters $\{W/\bar{d}_0,$ $\mu_{\rm w},$ $c^{\rm l}_0,$ $d^{\rm l}/d^{\rm s}\}$ specifies the geometry, loads, and initial conditions of a given case of vertical chute flow.  As a representative base case, we consider the parameter group $\{W/\bar{d}_0 = 60,\mu_{\rm w}=0.45,c^{\rm l}_0=0.5,d^{\rm l}/d^{\rm s}=1.5\}$. We then run the corresponding DEM simulation starting from the well-mixed initial configuration shown in Fig.~\ref{fig4_SegS}(a) and observe that after a simulation time of $\tilde{t} = t/( d^{\rm s}\sqrt{\rho_{\rm s}/P_{\rm w}} )= 4.3 \times 10^5$, the large, dark-gray grains segregate towards the regions near the walls where the shear-strain-rate is greatest, while the small, light-gray grains gather in bands just inside these regions, as shown in Fig.~\ref{fig4_SegS}(b). A well-mixed core persists along the center of the vertical chute where the shear-strain-rate is nearly zero. To obtain a more quantitative picture of the segregation process, we coarse-grain the concentration field $c^{\rm l}$ in both space and time and plot contours of the spatiotemporal evolution of $c^{\rm l}$ in Fig.~\ref{fig4_SegS}(c). The large-grain concentration field evolves quickly in time during flow initiation. Then, over longer times, the evolution becomes slower. Spatial profiles of the concentration and velocity fields at three snapshots in time--specifically, $\tilde{t} = t/( d^{\rm s}\sqrt{\rho_{\rm s}/P_{\rm w}} ) =4 \times 10^3$, $4 \times 10^4$ and $4 \times 10^5$ as indicated by the horizontal lines in Fig.~\ref{fig4_SegS}(c)--are plotted in Figs.~\ref{fig4_SegS}(d) and (e). These three snapshots correspond to early, medium, and late times with respect to the segregation process. The spatial $c^{\rm l}$ profiles shown in Fig.~\ref{fig4_SegS}(d) demonstrate the transition from a well-mixed state to a segregated state with large-grain-rich and small-grain-rich regions. In Fig.~\ref{fig4_SegS}(e), the normalized velocity fields $(v_{\rm cen} - v_z)\sqrt{\rho_{\rm s}/P_{\rm w}}$, relative to the velocity at the center of the chute, $v_{\rm cen} = v_z(x=0)$, show that the velocity field rapidly develops into a steady flow field, even while the segregation process is still ongoing, and the $c^{\rm l}$ field continues to evolve.

\begin{figure}[!t]
\begin{center}
\includegraphics{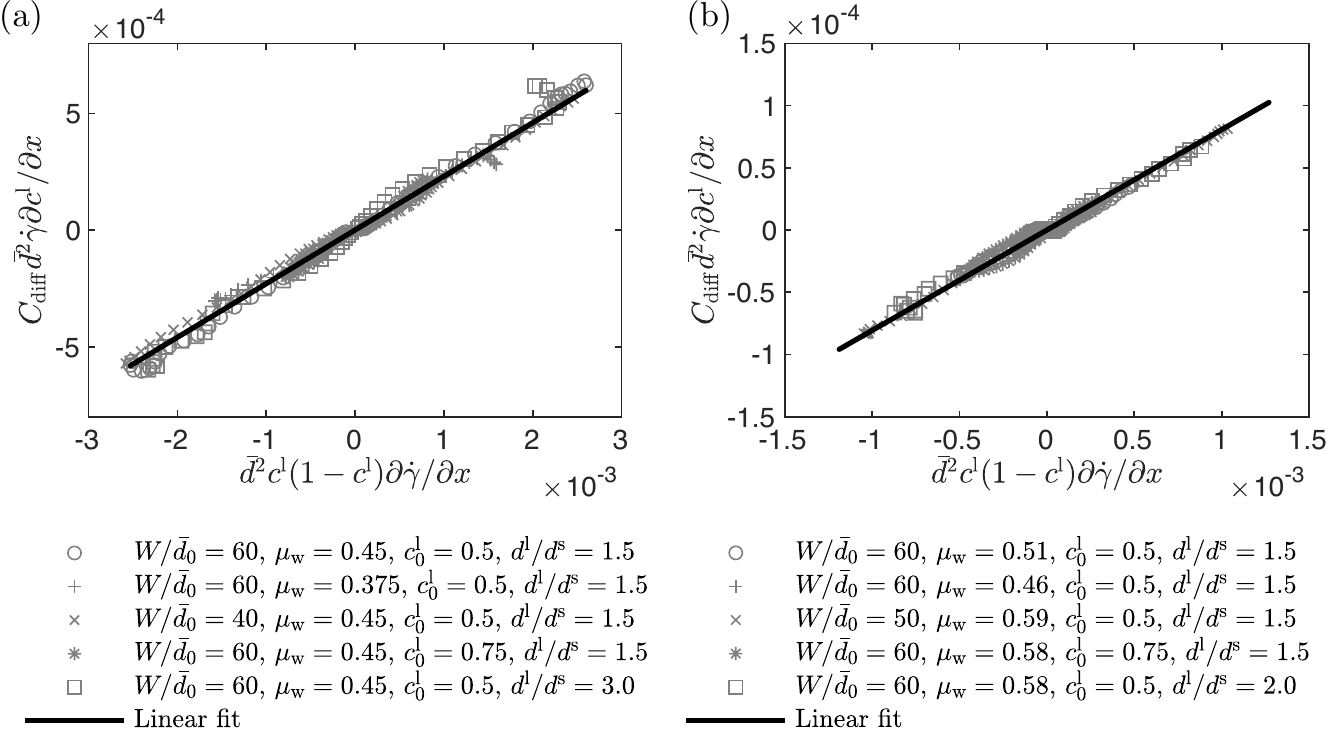}
\end{center}
\caption{Collapse of $C_{\rm diff}\bar{d}^2\dot\gamma({\partial c^{\rm l}}/{\partial x})$ versus $\bar{d}^2c^{\rm l}(1-c^{\rm l})({\partial \dot\gamma}/{\partial x})$ for several cases of vertical chute flow of (a) bidisperse disks and (b) bidisperse spheres. Symbols represent coarse-grained, quasi-steady DEM field data, and the solid lines are the best linear fits using (a) $C^{\rm S}_{\rm seg}=0.23$ for disks and (b) $C^{\rm S}_{\rm seg}=0.08$ for spheres.}\label{fig5_SegS}
\end{figure}

At long times, near the end of the simulated time window ($\tilde{t} = t/( d^{\rm s}\sqrt{\rho_{\rm s}/P_{\rm w}} )\gtrsim 3 \times 10^5$), the concentration field evolves very slowly, so that $D{c}^{\rm l}/Dt \approx 0$, and the state of segregation may be regarded as quasi-steady. Therefore, according to equations \eqref{mass_conserv_seg} and \eqref{flux_decomp_seg} and the no-flux boundary condition at the walls, the total flux is approximately zero ($w_i^{\rm l} = w_i^{\rm diff} + w_i^{\rm seg} \approx 0_i$) at each $x$-position, meaning that the segregation flux is approximately balanced by the diffusion flux in this quasi-steady flow regime. Then, using the expressions for the two fluxes, equations \eqref{diff_flux_seg} and \eqref{segS_flux_seg}, this observation implies that
\begin{equation}\label{flux_balance_VCF_seg}
C_{\rm diff}\bar{d}^2\dot\gamma\dfrac{\partial c^{\rm l}}{\partial x} \approx C^{\rm S}_{\rm seg}\bar{d}^2c^{\rm l}(1-c^{\rm l})\dfrac{\partial \dot\gamma}{\partial x}. 
\end{equation}
The field quantities appearing in this expression may be obtained by coarse-graining the DEM data in the quasi-steady flow regime. Therefore, since $C_{\rm diff}$ has been previously determined, \eqref{flux_balance_VCF_seg} may be used to determine the parameter $C^{\rm S}_{\rm seg}$ as follows. First, we acquire the field quantities $c^{\rm l}$ (and hence $\bar{d}$) and $v_z$ by spatially coarse-graining $152$ evenly-distributed snapshots in time in the quasi-steady regime ($\tilde{t} > 3 \times 10^5$).\footnote{Results are not particularly sensitive to the quasi-steady regime criterion, and we only provide this value as a guideline.} Then, we arithmetically average these fields in time, yielding fields that only depend on the spatial coordinate $x$, and take spatial gradients to obtain ${\partial c^{\rm l}}/{\partial x}$, $\dot\gamma = {\partial v_z}/{\partial x}$, and  ${\partial \dot\gamma}/{\partial x} = {\partial^2 v_z}/{\partial x^2}$.\footnote{Detailed descriptions of our coarse-graining techniques may be found in  Appendix~\ref{app_average}.} Next, as suggested by \eqref{flux_balance_VCF_seg}, we plot $C_{\rm diff}\bar{d}^2\dot\gamma({\partial c^{\rm l}}/{\partial x})$ versus $\bar{d}^2c^{\rm l}(1-c^{\rm l})({\partial \dot\gamma}/{\partial x})$ in Fig.~\ref{fig5_SegS}(a), in which each point represents a unique $x$-position. A linear relation is observed, supporting our choice for the form of the constitutive equation for the segregation flux \eqref{segS_flux_seg}. In order to obtain further evidence for this choice, we consider four additional cases: (1) a lower flow rate case $\{W/\bar{d}_0 = 60,\mu_{\rm w}=0.375,c^{\rm l}_0=0.5,d^{\rm l}/d^{\rm s}=1.5\}$; (2) a narrower  channel case $\{W/\bar{d}_0 = 40,\mu_{\rm w}=0.45,c^{\rm l}_0=0.5,d^{\rm l}/d^{\rm s}=1.5\}$; (3) a more large grains case $\{W/\bar{d}_0 = 60,\mu_{\rm w}=0.45,c^{\rm l}_0=0.75,d^{\rm l}/d^{\rm s}=1.5\}$; and  (4) a larger size ratio case $\{W/\bar{d}_0 = 60,\mu_{\rm w}=0.45,c^{\rm l}_0=0.5,d^{\rm l}/d^{\rm s}=3.0\}$. Coarse-graining the quasi-steady fields for each case and including the field data in Fig.~\ref{fig5_SegS}(a), we observe a strong linear collapse. Finally, the dimensionless material parameter $C_{\rm seg}^{\rm S}$ may be obtained from the slope of the linear relation in Fig.~\ref{fig5_SegS}(a) (indicated by the solid line). We determine the numerical value for disks to be $C^{\rm S}_{\rm seg}=0.23$ and note that this value indeed appears to be independent of the grain-size ratio $d^{\rm l}/d^{\rm s}$.

We carry out an analogous set of DEM simulations for dense, bidisperse mixtures of spheres to determine the value of $C^{\rm S}_{\rm seg}$ for spheres. The DEM setup for spheres is similar to that shown in Fig~\ref{fig4_SegS}(a) for disks with a domain size of length $L=20\bar{d}_0$ in the $z$-direction, width $W$ in the $x$-direction, which is varied in our DEM simulations, and out-of-plane thickness $H=10\bar{d}_0$ in the $y$-direction. Periodic boundary conditions are applied along both the $y$- and $z$-directions for spheres, and a constant compressive normal stress $\sigma_{xx} = -P_w$ is applied using the same feedback scheme as utilized for disks. In DEM simulations of dense flows of spheres, we observe normal stress differences, in which the normal stresses $\sigma_{yy}$ and $\sigma_{zz}$ are slightly different from the prescribed value of $\sigma_{xx}= -P_w$, which is a widely-reported feature of dense flows of spheres in the literature \citep[e.g.,][]{srivastava2021viscometric}. However, all normal stresses, and hence the pressure field, are spatially uniform, and segregation occurs only due to shear-strain-rate gradients. As for disks, the set of system parameters $\{W/\bar{d}_0,$ $\mu_{\rm w},$ $c^{\rm l}_0,$ $d^{\rm l}/d^{\rm s}\}$ specifies the geometry, loads, and initial conditions for a given case of vertical chute flow. Here, the dimensionless parameter $\mu_{\rm w} = \phi\rho_{\rm s}GW/2P_{\rm w}$ continues to control the total flow rate down the chute. We consider five different cases: (1) a base case $\{W/\bar{d}_0 = 60,\mu_{\rm w}=0.51,c^{\rm l}_0=0.5,d^{\rm l}/d^{\rm s}=1.5\}$; (2) a lower flow rate case $\{W/\bar{d}_0 = 60,\mu_{\rm w}=0.46,c^{\rm l}_0=0.5,d^{\rm l}/d^{\rm s}=1.5\}$; (3) a narrower channel and higher flow rate case $\{W/\bar{d}_0 = 50,\mu_{\rm w}=0.59,c^{\rm l}_0=0.5,d^{\rm l}/d^{\rm s}=1.5\}$; (4) a more large grains and higher flow rate case $\{W/\bar{d}_0 = 60,\mu_{\rm w}=0.58,c^{\rm l}_0=0.75,d^{\rm l}/d^{\rm s}=1.5\}$; and (5) a larger size ratio and higher flow rate case $\{W/\bar{d}_0 = 60,\mu_{\rm w}=0.58,c^{\rm l}_0=0.5,d^{\rm l}/d^{\rm s}=2.0\}$. Each case involves $\sim 20000$ flowing grains. After a sufficiently long simulation time, a quasi-steady state is attained in each case, implying the flux balance \eqref{flux_balance_VCF_seg}. The quasi-steady field quantities appearing in \eqref{flux_balance_VCF_seg} are extracted for 1000 snapshots in time using the coarse-graining techniques described in Appendix~\ref{app_average} and then arithmetically averaged in time. The calculated quasi-steady diffusion flux $C_{\rm diff}\bar{d}^2\dot\gamma({\partial c^{\rm l}}/{\partial x})$ at discrete $x$-positions for each of the five cases is plotted versus the calculated quantity $\bar{d}^2c^{\rm l}(1-c^{\rm l})({\partial \dot\gamma}/{\partial x})$ in Fig.~\ref{fig5_SegS}(b), and we observe a linear relation. Therefore, the form of the constitutive equation for the segregation flux \eqref{flux_balance_VCF_seg} is also applicable to dense, bidisperse systems of spheres. The numerical value of the dimensionless material parameter $C^{\rm S}_{\rm seg}$ for spheres is determined from the slope of the linear relation to be $C^{\rm S}_{\rm seg} = 0.08$.

\subsection{Annular shear flow}\label{subsec_segAS_seg}
The constitutive equation for the segregation flux \eqref{segS_flux_seg} and the fitted values of the material parameters should be general across different flow geometries. To test this for the case of disks, we apply the same process described in the preceding section for vertical chute flow to a different flow geometry--annular shear flow. In this flow geometry, flow is driven through motion of the boundary rather than by gravity, but as in vertical chute flow, the pressure field is spatially uniform, which eliminates hydrostatic pressure gradients so that only shear-strain-rate-gradient-driven size-segregation occurs.

\begin{figure}[!t]
\begin{center}
\includegraphics{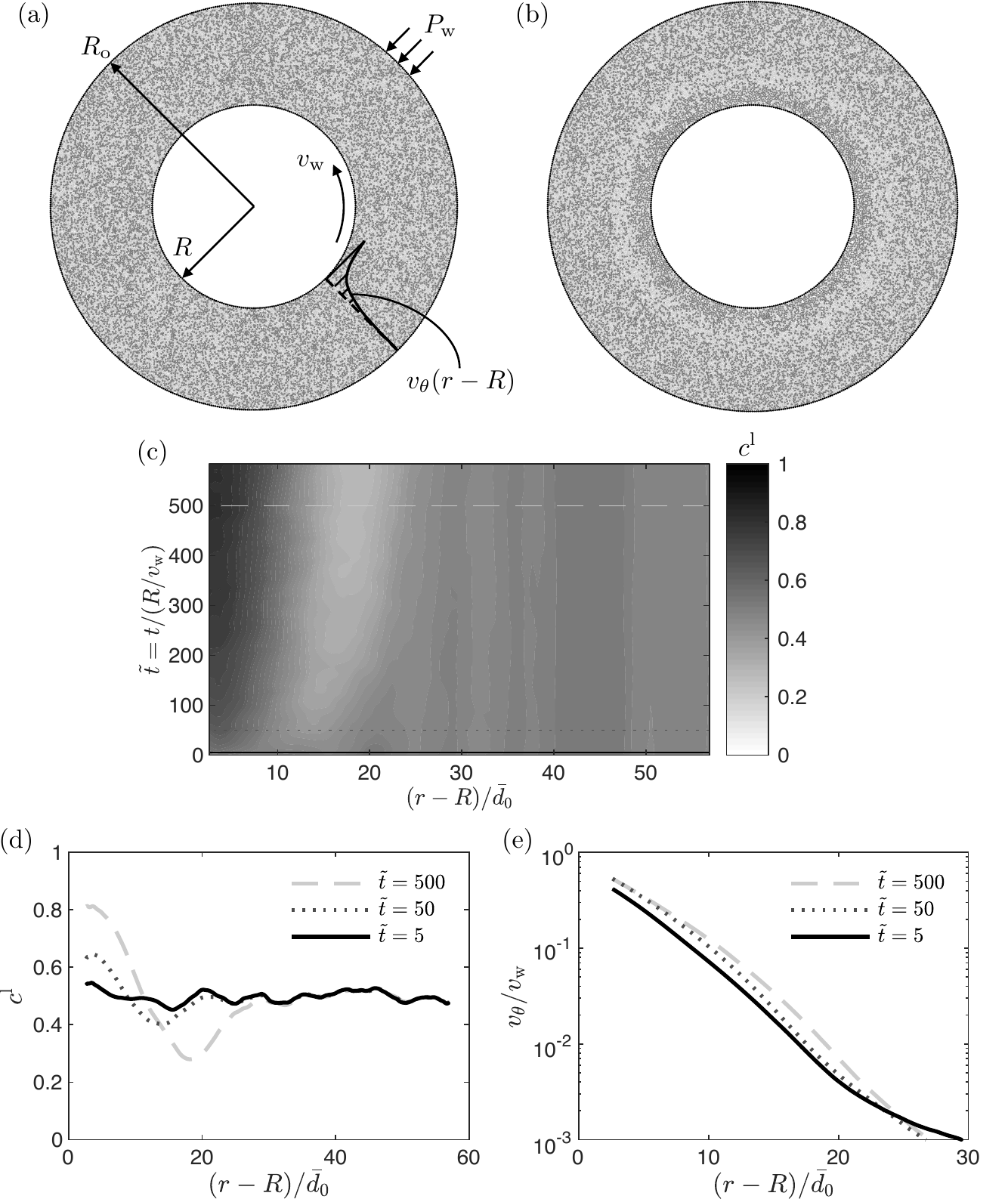}
\end{center}
\caption{(a) Initial well-mixed configuration for two-dimensional DEM simulation of bidisperse annular shear flow with $40108$ flowing grains. The inner wall radius is $R=60\bar{d}_0$, and the outer wall radius is $R_{\rm o}=2R$. The inner and outer walls consist of rings of glued large grains, denoted as black. (b) Segregated configuration after flowing for a total simulation time of $\tilde{t}=t/\left( R/v_{\rm w} \right)=584$. (c) Spatiotemporal evolution of the large grain concentration field. Spatial profiles of (d) the concentration field $c^{\rm l}$ and (e) the normalized circumferential velocity field ${v}_\theta/v_{\rm w}$ at three times ($\tilde{t}=5$, $50$ and $500$) as indicated by the horizontal lines in (c).}\label{fig6_SegS}
\end{figure}

Our DEM simulations of annular shear flow of a dense, bidisperse granular mixture of disks follow the procedures utilized in prior works in the literature for monodisperse, frictional disks \citep{koval2009,kamrin2012,kamrin2014,liu2018}. Consider a dense, bidisperse granular mixture in a two-dimensional annular shear cell with rough circular walls of inner radius $R$ and outer radius $R_{\rm o}$, as shown in Fig.~\ref{fig6_SegS}(a) for the case of $R=60\bar{d}_0$. The inner and outer walls consist of rings of glued large grains, denoted as black in Fig.~\ref{fig6_SegS}(a). The circumferential velocity of the inner wall is prescribed to be $v_{\rm w}$, and its radial position is fixed. The outer wall does not rotate, and its radius $R_{\rm o}$ fluctuates slightly so as to maintain a constant imposed compressive normal stress $P_{\rm w}$, utilizing the wall-position control method used throughout this work \citep{koval2009}. As in \citet{liu2018}, we simulate the full shear cell, as shown in Fig.~\ref{fig6_SegS}(a), instead of applying periodic boundary conditions along the circumferential direction to a slice \citep{koval2009,kamrin2012,kamrin2014}. In this flow geometry, all fields are axisymmetric (i.e., invariant along the $\theta$-direction), and the only non-zero component of the velocity field is the circumferential component $v_\theta$. Moreover, flow tends to localize near the inner wall with $v_\theta$ rapidly decaying with radial position, as qualitatively illustrated by the steady velocity field sketched in Fig.~\ref{fig6_SegS}(a). We choose the outer radius $R_{\rm o}$ to be sufficiently large so that it does not affect the resulting flow and segregation fields, and based on our experience, we take $R_{\rm o}=2R$. Therefore, the role of the outer wall is simply to apply a far-field pressure, and otherwise, it does not affect the flow and segregation fields. 

Regarding the stress field, in all of our DEM simulations of annular shear flow of bidisperse disks, we observe that the normal stresses are approximately equal, i.e., $\sigma_{rr}\approx\sigma_{\theta\theta}$, and spatially uniform, so that the force balance along the $r$-direction gives that the pressure field is $P(r) = -\sigma_{rr}(r) = P_{\rm w}$. Since the pressure field is spatially uniform, all segregation in annular shear flow is due to shear-strain-rate-gradients. The moment balance gives that the equivalent shear stress field is $\tau(r) = |\sigma_{r\theta}(r)| = |\sigma_{\theta r}(r)| = \tau_{\rm w}(R/r)^2$, where $\tau_{\rm w}$ is the inner-wall shear stress. It is important to note that $\tau_{\rm w}$ is not directly prescribed in our DEM simulations. Instead, the inner-wall velocity $v_{\rm w}$ is prescribed, and $\tau_{\rm w}$ arises as a result. The stress ratio field is then
\begin{equation}\label{AS_stress_seg}
\mu(r) = \mu_{\rm w}(R/r)^2,
\end{equation}
where $\mu_{\rm w}=\tau_{\rm w}/P_{\rm w}$ is the maximum value of $\mu$, occurring at the inner wall $(r = R)$.

As for vertical chute flow, there are four important dimensionless parameters that specify the geometry, loads, and initial conditions for a given case of annular shear flow of dense, bidisperse granular mixtures: (1) $R/\bar{d}_0$, the dimensionless inner-wall radius; (2) $\tilde{v}_{\rm w} = (v_{\rm w}/R)\sqrt{\pi\rho_{\rm s}\bar{d_0}^2/(4P_{\rm w})}$, the dimensionless inner-wall velocity, which determines $\tau_{\rm w}$ and hence $\mu_{\rm w}$; (3) $c^{\rm l}_0(r)$, the initial large-grain concentration field; and (4) $d^{\rm l}/d^{\rm s}$, the bidisperse grain-size ratio. We choose a representative base case of annular shear flow identified by the parameter set $\{R/\bar{d}_0 = 60,\tilde{v}_{\rm w}=0.01,c^{\rm l}_0=0.5,d^{\rm l}/d^{\rm s}=1.5\}$. The well-mixed initial configuration for the base-case DEM simulation is shown in Fig.~\ref{fig6_SegS}(a), and the segregated configuration after driving flow for a total simulation time of $\tilde{t}=t/\left( R/v_{\rm w} \right)=584$ is shown in Fig.~\ref{fig6_SegS}(b). The large, dark-gray grains segregate into a ring near the inner wall, while the small, light-gray grains form a band just outside this region. Outside of these bands, where the shear strain-rate is very small, the large and small grains remain well-mixed. Contours of the spatiotemporal evolution of the coarse-grained concentration field $c^{\rm l}$ are plotted in Figs.~\ref{fig6_SegS}(c), illustrating the time-evolution of $c^{\rm l}$ field. Spatial profiles of the concentration and velocity fields at three selected snapshots during the segregation dynamics ($\tilde{t}=t/\left( R/v_{\rm w} \right)=5, 50,$ and 500 as indicated by the dashed lines in Fig.~\ref{fig6_SegS}(c)) are shown in Figs.~\ref{fig6_SegS}(d) and (e). The radial $c^{\rm l}$ profiles in Fig.~\ref{fig6_SegS}(d) demonstrate the formation of large-grain-rich and small-grain-rich regions with a persistent well-mixed far-field. The normalized velocity fields in Fig.~\ref{fig6_SegS}(e) demonstrate that the flowing zone is localized near the inner wall with slow creeping flow observed far from the wall. As in the case of vertical chute flow, the velocity field quickly develops into a steady flow field, while the large grain concentration field $c^{\rm l}$ evolves over a longer time-scale before approaching a quasi-steady state near the end of our simulated time window. 

\begin{figure}[!t]
\begin{center}
\includegraphics{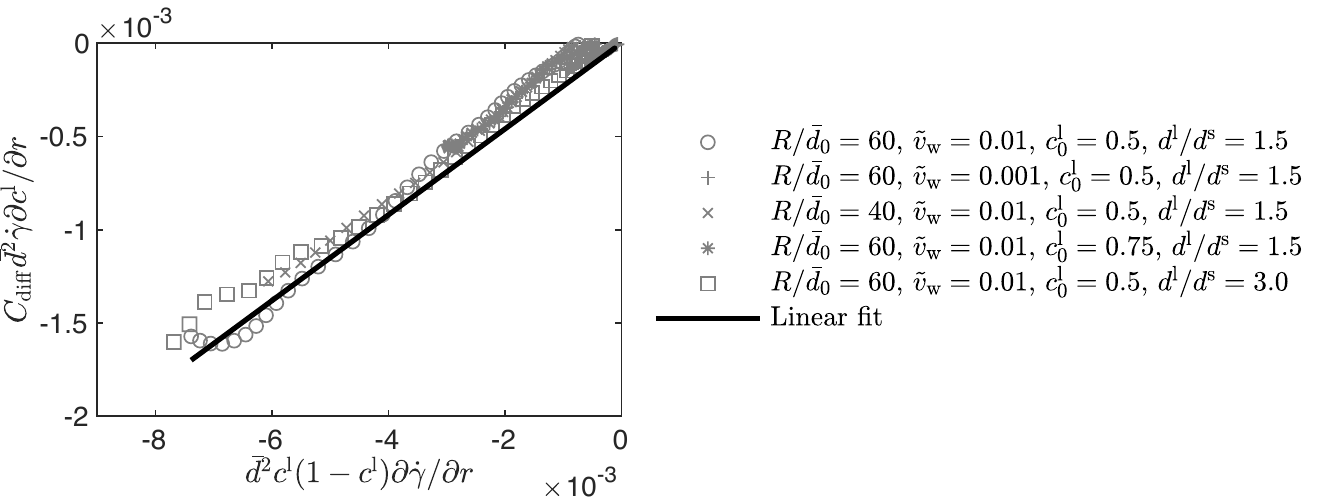}
\end{center}
\caption{Collapse of $C_{\rm diff}\bar{d}^2\dot\gamma({\partial c^{\rm l}}/{\partial r})$ versus $\bar{d}^2c^{\rm l}(1-c^{\rm l})({\partial \dot\gamma}/{\partial r})$ for several cases of annular shear flow of bidisperse disks. Symbols represent coarse-grained, quasi-steady DEM field data, and the solid line is the best linear fit using $C^{\rm S}_{\rm seg}=0.23$.}\label{fig7_SegS}
\end{figure}

Again, at long times, near the end of the simulated time window ($\tilde{t}=t/\left( R/v_{\rm w} \right)\gtrsim 500$), the concentration field evolves very slowly, which we identify as the quasi-steady regime. In this regime, the segregation and diffusion fluxes approximately balance at each $r$-position, implying that
\begin{equation}\label{flux_balance_AS_seg}
C_{\rm diff}\bar{d}^2\dot\gamma\dfrac{\partial c^{\rm l}}{\partial r} \approx C^{\rm S}_{\rm seg}\bar{d}^2c^{\rm l}(1-c^{\rm l})\dfrac{\partial \dot\gamma}{\partial r} .
\end{equation}
As for vertical chute flow, we spatially coarse-grain the DEM data to obtain the $c^{\rm l}$ and $v_\theta$ fields for $144$ evenly-distributed snapshots in time in the quasi-steady regime ($\tilde{t} \gtrsim 500$), which are arithmetically averaged in time to obtain the quasi-steady $c^{\rm l}(r)$ and $v_\theta(r)$ fields and then spatially differentiated to obtain the remaining field quantities in \eqref{flux_balance_AS_seg}. Next, we plot $C_{\rm diff}\bar{d}^2\dot\gamma({\partial c^{\rm l}}/{\partial r})$ versus $\bar{d}^2c^{\rm l}(1-c^{\rm l})({\partial \dot\gamma}/{\partial r})$ in Fig.~\ref{fig7_SegS} with each point representing a unique $r$-position. Finally, this process is repeated for four additional cases of annular shear flow: (1) a lower inner-wall velocity $\{R/\bar{d}_0 = 60,\tilde{v}_{\rm w}=0.001,c^{\rm l}_0=0.5,d^{\rm l}/d^{\rm s}=1.5\}$; (2) a smaller inner-wall radius $\{R/\bar{d}_0 = 40,\tilde{v}_{\rm w}=0.01,c^{\rm l}_0=0.5,d^{\rm l}/d^{\rm s}=1.5\}$; (3) more large grains $\{R/\bar{d}_0 = 60,\tilde{v}_{\rm w}=0.01,c^{\rm l}_0=0.75,d^{\rm l}/d^{\rm s}=1.5\}$; and  (4) a larger size ratio $\{R/\bar{d}_0 = 60,\tilde{v}_{\rm w}=0.01,c^{\rm l}_0=0.5,d^{\rm l}/d^{\rm s}=3.0\}$, and the coarse-grained, quasi-steady fields are included in the data plotted in Fig.~\ref{fig7_SegS}. Collectively, we observe a strong collapse to a linear relation. Crucially, the slope of the linear relation in Fig.~\ref{fig7_SegS} (indicated by the solid line) gives the same value for the dimensionless material parameter $C_{\rm seg}^{\rm S}$ obtained by fitting to vertical chute flow data, $C_{\rm seg}^{\rm S}=0.23$. This observation of agreement between the best fit values of $C_{\rm seg}^{\rm S}$ obtained using two different flow geometries provides support for our choice of the constitutive equation for the segregation flux \eqref{segS_flux_seg} and the fitted value of $C^{\rm S}_{\rm seg}$ for disks. Having established that the parameter $C^{\rm S}_{\rm seg}$ is independent of the flow geometry, driving conditions, and initial conditions, we henceforth regard $C^{\rm S}_{\rm seg}$ as a material parameter for a given dense granular system, analogous to how rheological material parameters such as $\mu_{\rm s}$ are regarded. Of course, the values of $C^{\rm S}_{\rm seg}$ determined above for disks and spheres likely depend on grain interaction properties, such as the inter-particle friction coefficient, but elucidating this dependence is beyond the scope of the present work. 

\section{Validation of the continuum model in the transient regime}\label{sec_shtrans_seg}

In the preceding section, we only used DEM data from the quasi-steady regime to test the constitutive equation for the shear-strain-rate-gradient-driven segregation flux \eqref{segS_flux_seg} and to determine the material parameter $C_{\rm seg}^{\rm S}$. In this section, we compare continuum model predictions of the transient evolution of segregation and flow fields to the DEM data for both vertical chute flow and annular shear flow as a validation test of the model. To obtain continuum model predictions in the transient regime, we couple the segregation dynamics equation \eqref{segS_model_seg} with the NGF model, \eqref{flowrule1D_seg} and \eqref{nonlocal_seg}, and a use fixed sets of material parameters for disks, 
\begin{equation}\label{mater_param_seg}
\{ \mu_{\rm s}=0.272, b=1.168, A=0.9, C_{\rm diff}=0.20, C^{\rm S}_{\rm seg}=0.23 \},
\end{equation}
and for spheres, 
\begin{equation}\label{mater_param_seg_spheres}
\{ \mu_{\rm s}=0.37, \mu_2=0.95, I_0=0.58, A=0.43, C_{\rm diff}=0.045, C^{\rm S}_{\rm seg}=0.08 \}.
\end{equation}

\subsection{Vertical chute flow}\label{subsec_transVC_seg}
First, we describe in detail how the continuum model is solved to obtain predictions for the transient evolution of segregation and flow fields for the case of vertical chute flow of disks. In vertical chute flow, the stress field may be straightforwardly deduced from a static force balance, giving that the pressure field is uniform, $P(x) = P_{\rm w}$, and that the stress ratio field $\mu(x)$ is given through \eqref{VCF_stress_seg}, and therefore, the balance of linear momentum \eqref{Chap_NGFmodel_eom} is satisfied and does not further enter the solution procedure. Continuum model predictions are obtained by numerically solving the remaining governing  equations using finite-differences. Summarizing the coupled boundary/initial-value problem for flow and segregation in the context of vertical chute flow, the unknown fields are the velocity field $v_z(x,t)$ and the accompanying strain-rate field $\dot\gamma(x,t) = \partial v_z/\partial x$, the granular fluidity field $g(x,t)$, and the large-grain concentration field $c^{\rm l}(x,t)$. The  governing equations are (1) the flow rule \eqref{flowrule1D_seg} 
\begin{equation}\label{flowrule1D_seg_sum}
\dot\gamma = g\mu,
\end{equation}
(2) the nonlocal rheology \eqref{nonlocal_seg} 
\begin{equation}\label{nonlocal_seg_sum}
g = g_{\rm loc}(\mu,P_{\rm w}) + \xi^2(\mu)\dfrac{\partial^2 g}{\partial x^2}
\end{equation}
with $g_{\rm loc}$ and $\xi$ given through \eqref{localg_bi_seg} and \eqref{cooperativity_bi_seg}$_1$, respectively, and (3) the segregation dynamics equation \eqref{segS_model_seg} 
\begin{equation}\label{segS_model_seg_sum}
    \dfrac{\partial c^{\rm l}}{\partial t} + \dfrac{\partial}{\partial x}\left(-C_{\rm diff}\bar{d}^2\dot\gamma\dfrac{\partial c^{\rm l}}{\partial x} + C^{\rm S}_{\rm seg}\bar{d}^2c^{\rm l}(1 - c^{\rm l})\dfrac{\partial \dot\gamma}{\partial x}\right) = 0,
\end{equation}
where $\bar{d} = c^{\rm l}d^{\rm l} + (1-c^{\rm l})d^{\rm s}$. 

Regarding boundary conditions, we impose Dirichlet fluidity boundary conditions at the walls (i.e., $g = g_{\rm loc}(\mu_{\rm w},P_{\rm w})$ at $x=\pm W/2$) as well as no flux boundary conditions at the walls (i.e., $w_x^{\rm l} = -C_{\rm diff} \bar{d}^2 \dot\gamma (\partial c^{\rm l}/\partial x) + C^{\rm S}_{\rm seg} \bar{d}^2 c^{\rm l}(1 - c^{\rm l})(\partial \dot\gamma/\partial x) = 0$ at $x=\pm W/2$). Due to the time derivative in \eqref{segS_model_seg_sum}, an initial condition for the concentration field $c^{\rm l}_0(x) = c^{\rm l}(x,t=0)$ is required. In order to account for the concentration fluctuations inherent in the initial state, we obtain the coarse-grained $c^{\rm l}$-field from the initial DEM configuration for each case and utilize this field as the initial condition field $c_0^{\rm l}(x)$ in each of the respective continuum simulations.

Then, for a given case identified through a set of input parameters $\{W/\bar{d}_0,$ $\mu_{\rm w},$ $c^{\rm l}_0(x),$ $d^{\rm l}/d^{\rm s}\}$, we obtain numerical predictions of the continuum model utilizing finite differences as follows. First, at a given point in time, the concentration field $c^{\rm l}(x)$ is known, allowing the average grain-size field to be calculated through $\bar{d} = c^{\rm l}d^{\rm l} + (1-c^{\rm l})d^{\rm s}$. Using the stress ratio field $\mu(x)$ for vertical chute flow \eqref{VCF_stress_seg}, the local fluidity  $g_{\rm loc}(\mu,P)$ and the cooperativity length $\xi(\mu)$ (equations \eqref{localg_bi_seg} and \eqref{cooperativity_bi_seg}$_1$)  may be calculated at each spatial grid point. Then, the nonlocal rheology \eqref{nonlocal_seg_sum} may be used to solve for the fluidity field $g(x)$ at the current step, using central differences in space. The strain-rate field follows using \eqref{flowrule1D_seg_sum}, which may be integrated to obtain the velocity field $v_z(x)$. Next, \eqref{segS_model_seg_sum} is used to determine the concentration field at the next time step utilizing the forward Euler method and central differences in space with one modification--the spatial derivatives of $c^{\rm l}$ appearing in the diffusion flux term in \eqref{segS_model_seg_sum} are treated implicitly in order to improve numerical stability. This completes one time step, and this process is repeated to step forward in time and calculate the transient evolution of the concentration and flow fields, $c^{\rm l}(x,t)$ and $v_z(x,t)$. In our finite-difference calculations, we utilize a fine spatial resolution of $\Delta x \ll \bar{d}_0$, and we have verified that the time-step is sufficiently small in order to ensure stable, accurate results.

\begin{figure}[!t]
\begin{center}
\includegraphics{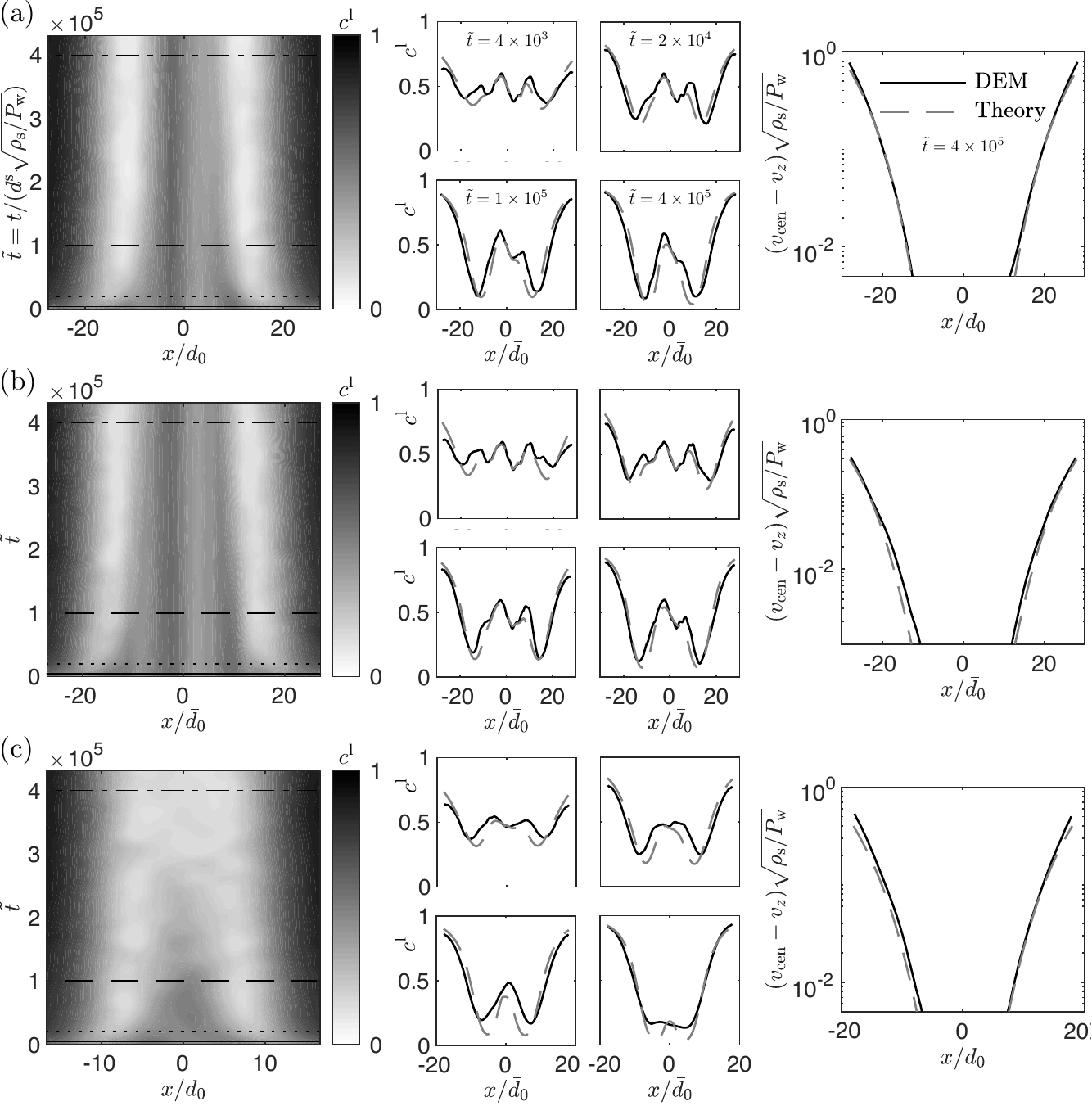}
\end{center}
\caption{Comparisons of continuum model predictions with corresponding DEM simulation results for the transient evolution of the segregation dynamics for three cases of vertical chute flow of disks: (a) Base case $\{W/\bar{d}_0 = 60,\mu_{\rm w}=0.45,c^{\rm l}_0=0.5,d^{\rm l}/d^{\rm s}=1.5\}$; (b) Lower flow rate case $\{W/\bar{d}_0 = 60,\mu_{\rm w}=0.375,c^{\rm l}_0=0.5,d^{\rm l}/d^{\rm s}=1.5\}$; and (c) Narrower chute width case $\{W/\bar{d}_0 = 40,\mu_{\rm w}=0.45,c^{\rm l}_0=0.5,d^{\rm l}/d^{\rm s}=1.5\}$. Additional cases are shown in Fig.~\ref{fig8_SegS_part2}. For each case, the first column shows spatiotemporal contours of the evolution of $c^{\rm l}$ measured in the DEM simulations. The second column shows comparisons of the DEM simulations (solid black lines) and continuum model predictions (dashed gray lines) of the $c^{\rm l}$ field at four time snapshots representing different stages of the segregation process: $\tilde{t}=4 \times 10^3$, $2 \times 10^4$, $1 \times 10^5$, and $4 \times 10^5$ in the sequence of top left, top right, bottom left, bottom right. The third column shows comparisons of the quasi-steady, normalized velocity profiles at $\tilde{t}=4 \times 10^5$ from DEM simulations and continuum model predictions.} \label{fig8_SegS}
\end{figure}

\begin{figure}[!t]
\begin{center}
\includegraphics{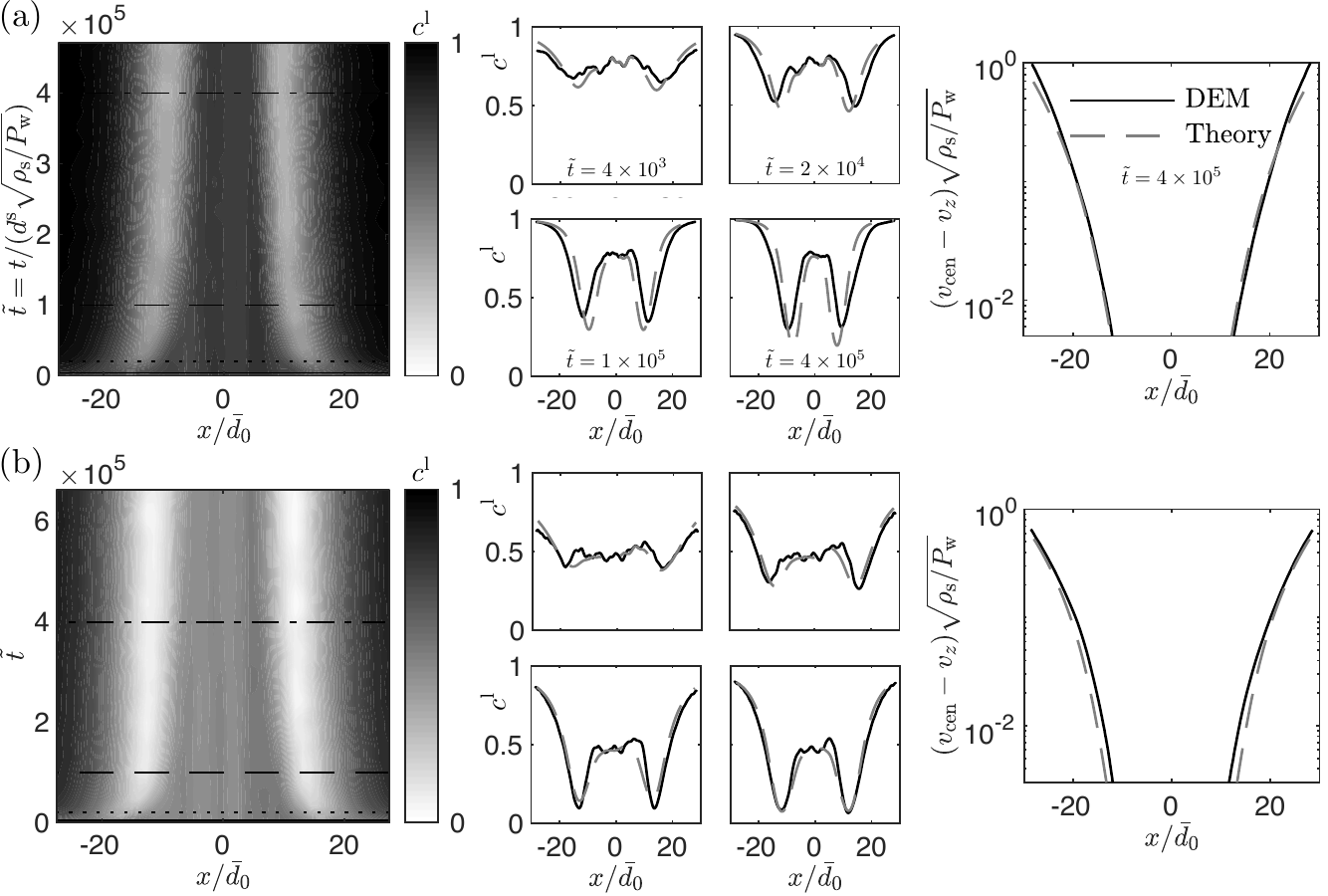}
\end{center}
\caption{Comparisons of continuum model predictions with corresponding DEM simulation results for the transient evolution of the segregation dynamics for two cases of vertical chute flow of disks: (a) More large grains case $\{W/\bar{d}_0 = 60,\mu_{\rm w}=0.45,c^{\rm l}_0=0.75,d^{\rm l}/d^{\rm s}=1.5\}$ and (b) Larger size ratio case $\{W/\bar{d}_0 = 60,\mu_{\rm w}=0.45,c^{\rm l}_0=0.5,d^{\rm l}/d^{\rm s}=3.0\}$. Additional cases are shown in Fig.~\ref{fig8_SegS}. Results are organized as described in the caption of Fig.~\ref{fig8_SegS}.}\label{fig8_SegS_part2}
\end{figure}

\begin{figure}[!t]
\begin{center}
\includegraphics{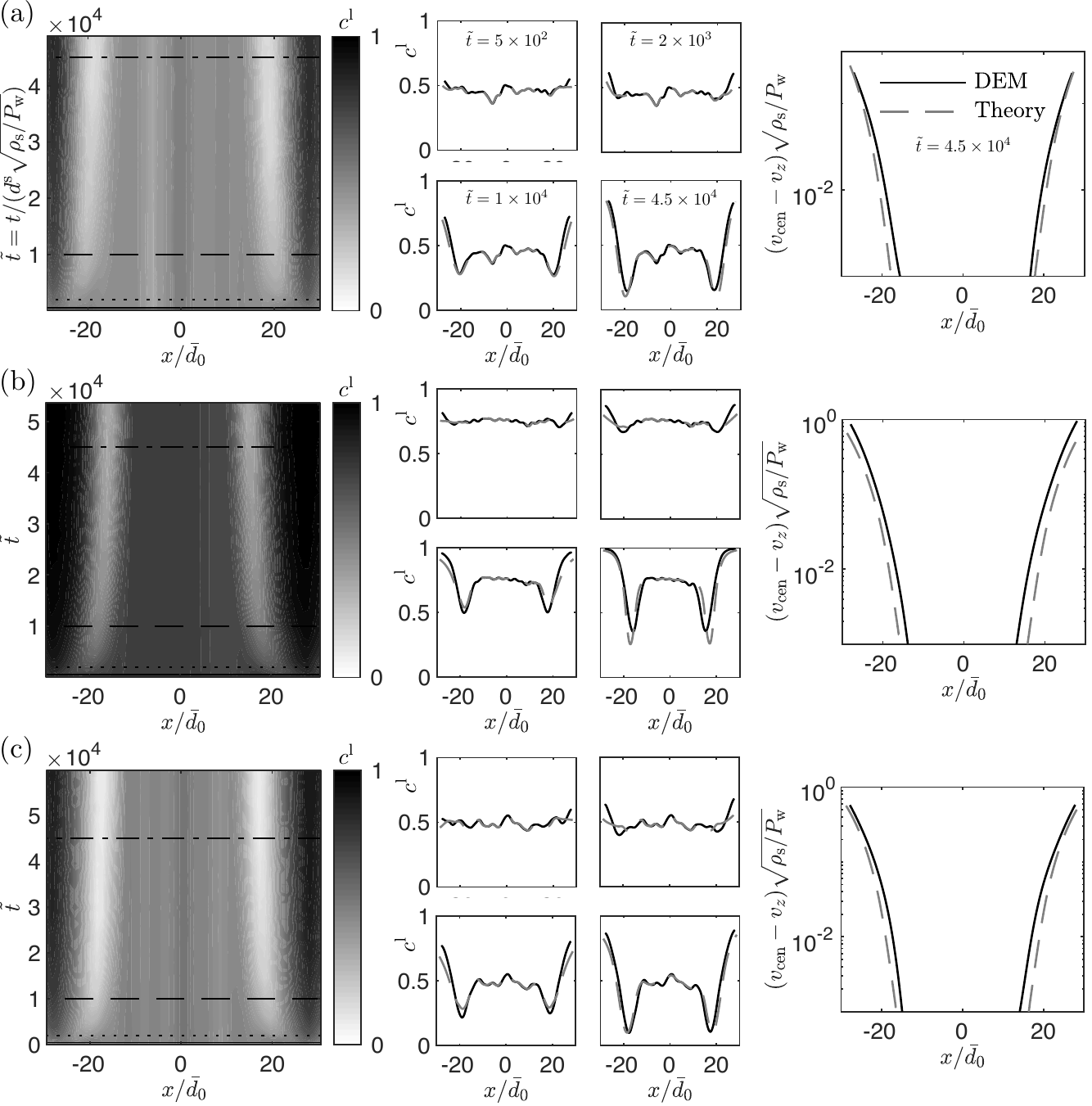}
\end{center}
\caption{Comparisons of continuum model predictions with corresponding DEM simulation results for the transient evolution of the segregation dynamics for three cases of vertical chute flow of spheres: (a) Base case $\{W/\bar{d}_0 = 60,\,\mu_{\rm w}=0.51,\,c^{\rm l}_0=0.5,\,d^{\rm l}/d^{\rm s}=1.5\}$; (b) More large grains and higher flow rate case $\{W/\bar{d}_0 = 60,\mu_{\rm w}=0.58,c^{\rm l}_0=0.75,d^{\rm l}/d^{\rm s}=1.5\}$; and (c) Larger grain-size ratio and higher flow rate case $\{W/\bar{d}_0 = 60,\,\mu_{\rm w}=0.58,\,c^{\rm l}_0=0.5,\,d^{\rm l}/d^{\rm s}=2.0\}$. Results are organized as described in the caption of Fig.~\ref{fig8_SegS}.} \label{fig_SegS_spheres}
\end{figure}

We compare predictions of the continuum model against DEM data for all five cases of vertical chute flow considered in Section \ref{subsec_segVCF_seg} in order to test the generality of the model. Figures \ref{fig8_SegS} and \ref{fig8_SegS_part2} summarize the comparisons for these five cases. The first column of Figs.~\ref{fig8_SegS} and \ref{fig8_SegS_part2} shows the spatiotemporal contours of the evolution of the $c^{\rm l}$ field measured in the DEM simulations for each case, and the second column shows comparisons of the DEM simulations (solid black lines) and the continuum predictions (dashed gray lines) for the $c^{\rm l}$ field at four snapshots in time ($\tilde{t} = t/\left( d^{\rm s}\sqrt{\rho_{\rm s}/P_{\rm w}} \right) =4 \times 10^3$, $2 \times 10^4$, $1 \times 10^5$ and $4 \times 10^5$) indicated by the horizontal lines in the first column of Figs.~\ref{fig8_SegS} and \ref{fig8_SegS_part2}. Based on Figs.~\ref{fig8_SegS} and \ref{fig8_SegS_part2}, the coupled model generally does a good job capturing the salient features of the evolution of the $c^{\rm l}$ field across all cases. For instance, for the narrower chute case shown in Fig.~\ref{fig8_SegS}(c), the segregation process nearly completes within the simulated time window with the mixed core along the center of the chute nearly disappearing, and the continuum model prediction captures this observation well.

Regarding flow fields, comparisons of the quasi-steady, normalized velocity fields at $\tilde{t} = 4 \times 10^5$ from the DEM simulations and the continuum model predictions are shown in the third column of Fig.~\ref{fig8_SegS}. Since the velocity field evolves minimally during the segregation process, only the flow field at long time is shown. We note that the velocity field is very well-predicted in all cases, including the creeping regions far from the wall. This favorable comparison provides support of our generalization of the NGF model to bidisperse granular systems discussed in Section \ref{subsec_rheo_seg}--in particular, the choices to use the average grain size $\bar{d}$ in the expression for the cooperativity length \eqref{cooperativity_bi_seg} and to continue to use the numerical value for the nonlocal amplitude determined for monodisperse systems, $A=0.9$, without refitting. 

Next, we make comparisons between continuum model predictions and DEM data for vertical chute flow of bidisperse spheres. The governing equations and boundary conditions are same as those used above for bidisperse disks. That is, the fluidity field is governed by the nonlocal rheology \eqref{nonlocal_seg_sum} with Dirichlet fluidity boundary conditions at the walls ($g=g_{\rm loc}(\mu_{\rm w}, P_{\rm w})$ at $x=\pm W/2$), and the concentration field is governed by the segregation dynamics equation \eqref{segS_model_seg_sum} with no flux boundary conditions at the walls ($w_x^{\rm l}=0$ at $x=\pm W/2$). The only difference from the process described above for bidisperse disks is that the values $P_{\rm w}$ and $\mu_{\rm w}$ used in the continuum simulations are obtained from the coarse-grained stress fields in the DEM data for each case, rather than based on the nominal value of the compressive wall stress $P_{\rm w}$ applied in the DEM simulation. This is done to account for the normal stress differences that arise for spheres, which slightly affect the predicted velocity fields and hence the consequent concentration fields. We consider three different cases: (1) the base case $\{W/\bar{d}_0 = 60,\,\mu_{\rm w}=0.51,\,c^{\rm l}_0=0.5,\,d^{\rm l}/d^{\rm s}=1.5\}$, (2) the more large grains and higher flow rate case $\{W/\bar{d}_0 = 60,\,\mu_{\rm w}=0.58,\,c^{\rm l}_0=0.75,\,d^{\rm l}/d^{\rm s}=1.5\}$, and (3) the larger size ratio and higher flow rate case $\{W/\bar{d}_0 = 60,\,\mu_{\rm w}=0.58,\,c^l_0=0.5,\,d^l/d^s=2.0\}$, which are shown in Figs.~\ref{fig_SegS_spheres}(a), (b), and (c), respectively. The leftmost column of Fig.~\ref{fig_SegS_spheres} shows the spatiotemporal contours of the evolution of the $c^{\rm l}$ field from the DEM data. The middle column shows comparisons between the DEM simulations (solid black lines) and the continuum model predictions (dashed gray lines)  for the $c^{\rm l}$ fields at four time snapshots in time ($\tilde{t} = t/\left( d^{\rm s}\sqrt{\rho_{\rm s}/P_{\rm w}} \right) =5 \times 10^2$, $2 \times 10^3$, $1 \times 10^4$ and $4.5 \times 10^4$) indicated by the horizontal lines in the first column of Fig.~\ref{fig_SegS_spheres}. Lastly, comparisons of the quasi-steady, normalized velocity fields at $\tilde{t} = 4.5 \times 10^4$ from the DEM simulations and the continuum model predictions are shown in the third column of Fig.~\ref{fig_SegS_spheres}. The continuum model is able to capture the decaying velocity field quite well in all cases, and therefore, our choice to continue using the value of the nonlocal amplitude estimated for monodisperse systems, $A=0.43$, without readjustment works well for spheres. 
Overall, the coupled continuum model is capable of quantitatively predicting both the flow fields and the transient evolution of the segregation dynamics in vertical chute flow of bidisperse spheres. 

\begin{figure}[!t]
\begin{center}
\includegraphics{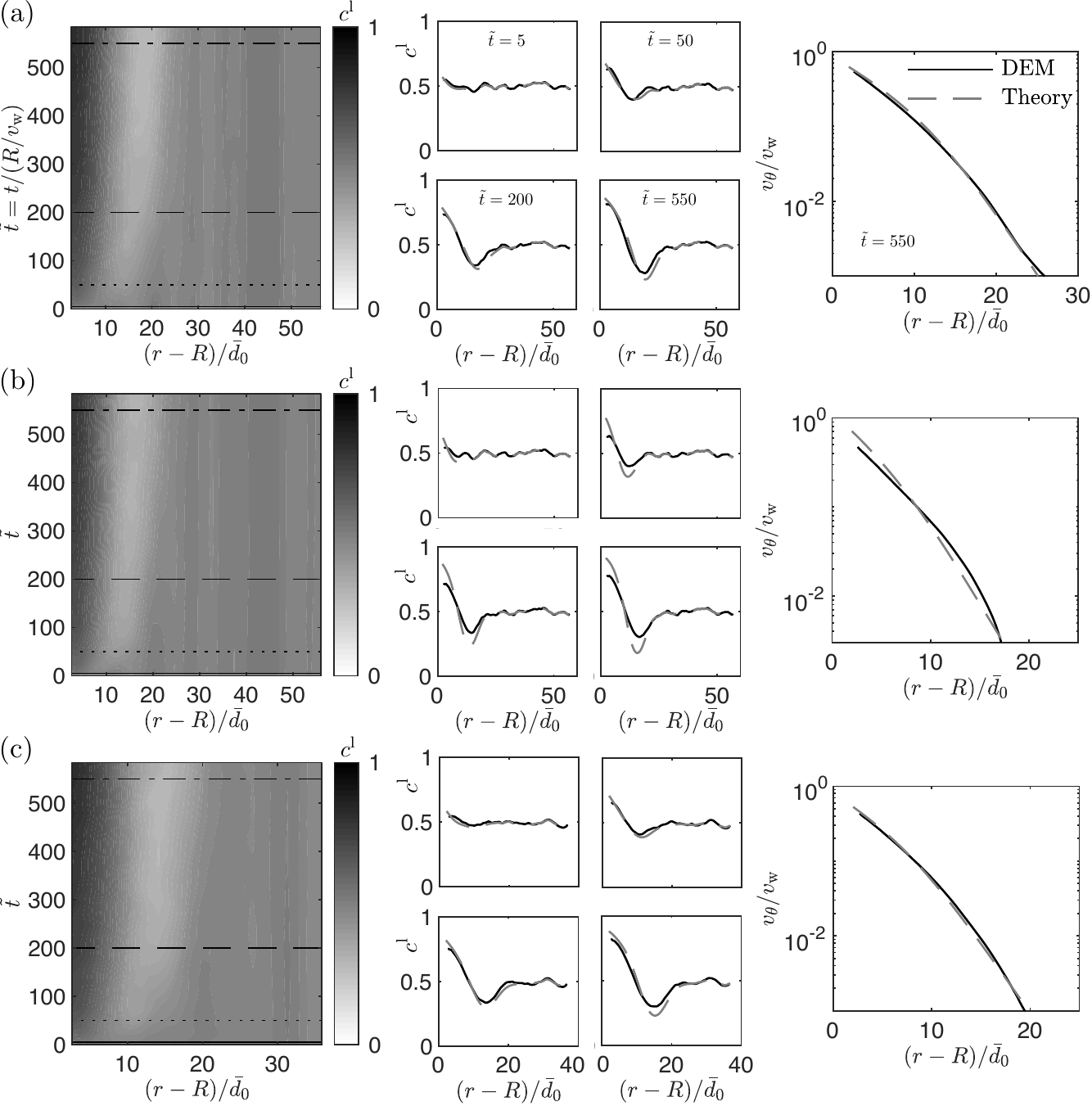}
\end{center}
\caption{Comparisons of the continuum model predictions with corresponding DEM simulation results for the transient evolution of the segregation dynamics for three cases of annular shear flow of disks: (a) Base case $\{R/\bar{d}_0 = 60,\tilde{v}_{\rm w}=0.01,c^{\rm l}_0=0.5,d^{\rm l}/d^{\rm s}=1.5\}$; (b) Lower inner-wall velocity case $\{R/\bar{d}_0 = 60,\tilde{v}_{\rm w}=0.001,c^{\rm l}_0=0.5,d^{\rm l}/d^{\rm s}=1.5\}$; and (c) Smaller annular shear cell case $\{R/\bar{d}_0 = 40,\tilde{v}_{\rm w}=0.01,c^{\rm l}_0=0.5,d^{\rm l}/d^{\rm s}=1.5\}$. Additional cases are shown in Fig.~\ref{fig9_SegS_part2}. For each case, the first column shows spatiotemporal contours of the evolution of $c^{\rm l}$ measured in the DEM simulations. The second column shows comparisons of the DEM simulations (solid black lines) and continuum model predictions (dashed gray lines) of the $c^{\rm l}$ field at four time snapshots representing different stages of the segregation process: $\tilde{t}=5$, $50$, $200$, and $550$ in the sequence of top left, top right, bottom left, bottom right. The third column shows comparisons of the quasi-steady, normalized velocity profiles at $\tilde{t}=550$ from DEM simulations and continuum model predictions.}\label{fig9_SegS}
\end{figure}

\begin{figure}[!t]
\begin{center}
\includegraphics{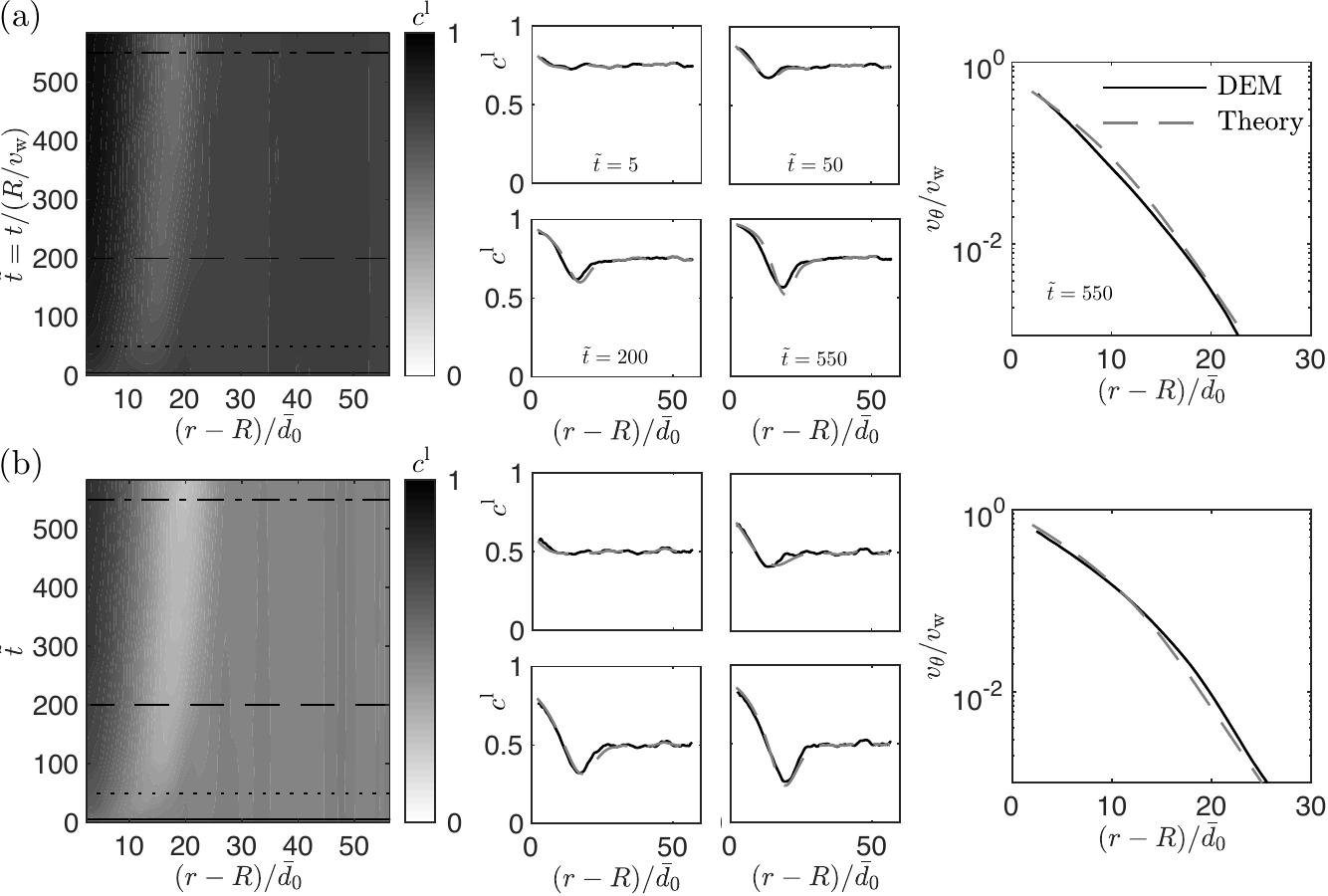}
\end{center}
\caption{Comparisons of the continuum model predictions with corresponding DEM simulation results for the transient evolution of the segregation dynamics for two cases of annular shear flow of disks: (a) More large grains case $\{R/\bar{d}_0 = 60,\tilde{v}_{\rm w}=0.01,c^{\rm l}_0=0.75,d^{\rm l}/d^{\rm s}=1.5\}$ and  (b) Larger size ratio case $\{R/\bar{d}_0 = 60,\tilde{v}_{\rm w}=0.01,c^{\rm l}_0=0.5,d^{\rm l}/d^{\rm s}=3.0\}$. Additional cases are shown in Fig.~\ref{fig9_SegS}. Results are organized as described in the caption of Fig.~\ref{fig9_SegS}.}\label{fig9_SegS_part2}
\end{figure}

\subsection{Annular shear flow}\label{subsec_transAS_seg}
We utilize an analogous process to obtain continuum model predictions for the transient evolution of concentration and flow fields in annular shear flow of disks. In annular shear, based on the static force and moment balances, the pressure field is uniform, $P(r)=P_{\rm w}$, and the stress ratio field is given by \eqref{AS_stress_seg}. The governing equations \eqref{nonlocal_seg_sum} and \eqref{segS_model_seg_sum} are modified to appropriately account for the divergence and Laplacian operators in cylindrical coordinates. Also, since $v_{\rm w}$, not $\mu_{\rm w}$, is specified in our DEM simulations of annular shear, while $\mu_{\rm w}$ is specified in our continuum simulations, we iteratively adjust the value of $\mu_{\rm w}$ input into our continuum simulations in order to achieve the target value of $v_{\rm w}$ in the predicted quasi-steady flow field. Otherwise, our process for obtaining numerical predictions from the continuum model is the same. Dirichlet fluidity boundary conditions and no flux boundary conditions are imposed at the walls, and the initial concentration field is extracted from the initial DEM configuration for each case.

Then, we compare continuum model predictions against DEM data for the five cases of annular shear flow discussed in Section~\ref{subsec_segAS_seg} in order to further validate the model. Figures~\ref{fig9_SegS} and \ref{fig9_SegS_part2} summarize the comparisons for these fives cases and are organized in the same manner as Figs.~\ref{fig8_SegS} and \ref{fig8_SegS_part2}. Again, the coupled, continuum model does a good job capturing the segregation dynamics and its dependence on the input parameters. Moreover, the quasi-steady velocity fields are well-predicted by the NGF model in all cases, including the creeping region far from the inner wall. We reiterate that all continuum model predictions are obtained using the same set of material parameters for disks \eqref{mater_param_seg}.

\section{Discussion and Conclusion}\label{sec_concl_seg}
In this paper, we studied coupled size-segregation and flow in dense, bidisperse granular systems of disks and spheres and developed a phenomenological continuum model that captures the simultaneous evolution of both segregation and flow fields. We focused on the shear-strain-rate-gradient-driven size-segregation mechanism in two configurations in which the pressure field is uniform--vertical chute flow and annular shear flow--and based on observations from DEM simulations, we proposed a phenomenological constitutive equation for the shear-strain-rate-gradient-driven flux. When combined with a standard model for granular diffusion, the segregation model involves two dimensionless parameters $\{C_{\rm diff},C^{\rm S}_{\rm seg}\}$, which multiply the two fluxes appearing in the model--the diffusion and shear-strain-rate-gradient-driven fluxes, respectively. By coupling the segregation model with the NGF model adapted to bidisperse systems, we may  quantitatively predict both the flow fields and the segregation dynamics for dense flows of bidisperse disks and spheres for two distinct flow geometries and under a number of different flow conditions. 

\begin{figure}[!t]
\begin{center}
\includegraphics{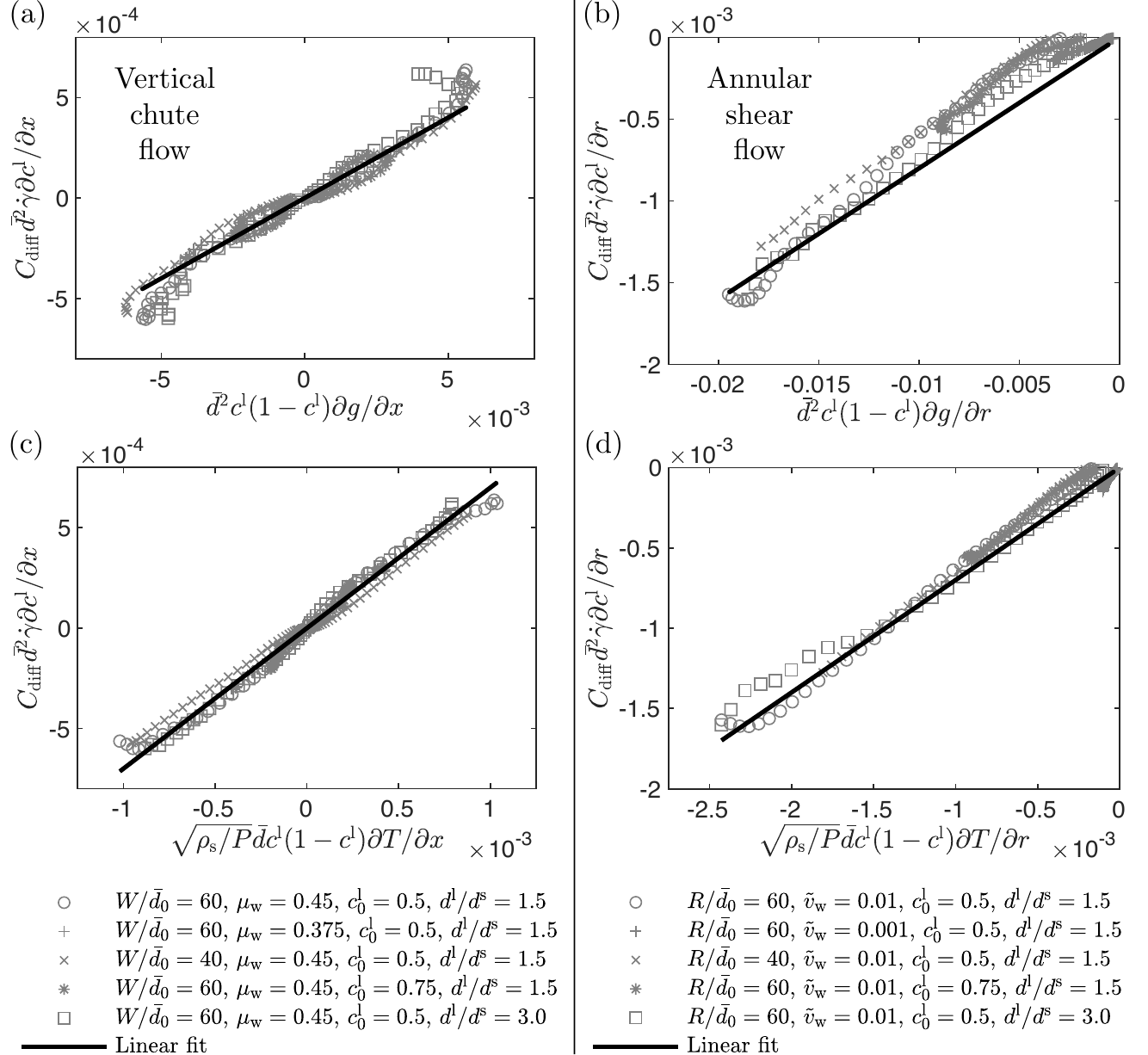}
\end{center}
\caption{(a) Collapse of $C_{\rm diff}\bar{d}^2\dot\gamma({\partial c^{\rm l}}/{\partial x})$ versus $\bar{d}^2c^{\rm l}(1-c^{\rm l})({\partial g}/{\partial x})$ for several cases of vertical chute flow and (b) collapse $C_{\rm diff}\bar{d}^2\dot\gamma({\partial c^{\rm l}}/{\partial r})$ versus $\bar{d}^2c^{\rm l}(1-c^{\rm l})({\partial g}/{\partial r})$ for several cases of annular shear flow of bidisperse disks. (c) Collapse of $C_{\rm diff}\bar{d}^2\dot\gamma({\partial c^{\rm l}}/{\partial x})$ versus $\sqrt{\rho_{\rm s}/P}\bar{d}c^{\rm l}(1-c^{\rm l})({\partial T}/{\partial x})$ for several cases of vertical chute flow and (d) collapse of $C_{\rm diff}\bar{d}^2\dot\gamma({\partial c^{\rm l}}/{\partial x})$ versus $\sqrt{\rho_{\rm s}/P}\bar{d}c^{\rm l}(1-c^{\rm l})({\partial T}/{\partial x})$ for several cases of annular shear flow of bidisperse disks. Symbols represent coarse-grained, quasi-steady DEM field data, and the solid lines represent the best linear fit using $C^{\rm S}_{\rm seg}=0.08$ for (a) and (b) and $C^{\rm S}_{\rm seg}=0.7$ for (c) and (d).}\label{fig10_SegS}
\end{figure}

Size-segregation in granular materials is a complex and rich problem, so there remain many avenues for model improvement and unresolved research questions to be answered. One important question relates to the constitutive equation for the shear-strain-rate-gradient-driven segregation flux \eqref{segS_flux_seg}. Although our use of a constitutive equation driven by gradients in $\dot\gamma$ does a good job capturing the DEM data, there are other theories in the literature based on gradients of other field quantities. In particular, Hill and coworkers \citep{fan11b,hill2014} have proposed that gradients in the kinetic stress, which is related to velocity fluctuations and hence the granular temperature,  drive segregation. Since \citet{zhang2017} have established a connection between velocity fluctuations and the granular fluidity $g$, it is possible to propose other forms for the constitutive equation for $w_i^{\rm seg}$ based on gradients in $g$. For example, instead of \eqref{segS_flux_seg}, consider the following form for the segregation flux:
\begin{equation}\label{segS_flux_g_seg}
w_i^{\rm seg} = C^{\rm S}_{\rm seg} \bar{d}^2c^{\rm l}(1-c^{\rm l})\dfrac{\partial g}{\partial x_i}.
\end{equation}
Then, applying the quasi-steady flux balance condition,
\begin{equation}\label{flux_balance_g_seg}
C_{\rm diff}\bar{d}^2\dot\gamma\dfrac{\partial c^{\rm l}}{\partial x_i}\approx C^{\rm S}_{\rm seg} \bar{d}^2c^{\rm l}(1-c^{\rm l})\dfrac{\partial g}{\partial x_i},
\end{equation}
to the quasi-steady DEM data for vertical chute flow and annular shear flow of disks, we obtain the collapses shown in Figs.~\ref{fig10_SegS}(a) and (b), respectively.\footnote{The coarse-grained values of $g$ and its gradient are obtained using the coarse-grained values of $\dot\gamma$ and its gradient, calculated as described in Appendix~\ref{app_average}, along with $\mu$ and its gradient, calculated using \eqref{VCF_stress_seg} and not by coarse-graining.} The solid lines represent the best linear fit using $C^{\rm S}_{\rm seg} = 0.08$. The collapses are reasonable but not as strong as those shown in Figs.~\ref{fig5_SegS}(a) and \ref{fig7_SegS} for a segregation flux based on gradients in the shear-strain-rate, leading us to choose to work with the constitutive equation \eqref{segS_flux_seg} on pragmatic grounds. 

Additionally, we have tested a possible constitutive equation for the segregation flux driven by gradients in the granular temperature, defined as $T=(\delta v)^2$, where $\delta v$ is the velocity fluctuation.\footnote{The definitions of the granular temperature $T$ and the velocity fluctuation $\delta v$ as well as the coarse-graining method used to obtain these quantities follow \citet{zhang2017}. They are treated as field quantities just as $\dot\gamma$. Following the coarse-graining process described in \eqref{app_average}, the instantaneous $T$ field at a bin is the grain-area-weighted summation of the square of the difference of the grain instantaneous velocity vector and the coarse-grained instantaneous velocity field at that grain center, for all grains that are intersected by the bin.} Then, consider the following form for the segregation flux: 
\begin{equation}\label{segS_flux_T_seg}
w_i^{\rm seg} = C^{\rm S}_{\rm seg}  \sqrt{\rho_{\rm s}/P} \bar{d} c^{\rm l}(1-c^{\rm l})\dfrac{\partial T}{\partial x_i},
\end{equation}
where the inertial time $\sqrt{\rho_{\rm s}/P} \bar{d}$ is included in the prefactor for dimensional reasons. As above, upon applying the quasi-steady flux balance condition, 
\begin{equation}\label{flux_balance_T_seg}
C_{\rm diff}\bar{d}^2\dot\gamma\dfrac{\partial c^{\rm l}}{\partial x_i}\approx C^{\rm S}_{\rm seg} \sqrt{\rho_{\rm s}/P} \bar{d} c^{\rm l}(1-c^{\rm l})\dfrac{\partial T}{\partial x_i},
\end{equation}
to the quasi-steady DEM data, Figs.~\ref{fig10_SegS}(c) and (d) show the collapses for granular-temperature-gradient-driven segregation for vertical chute flow and annular shear flow of disks, respectively. In this case, the solid lines represent the best linear fit using $C^{\rm S}_{\rm seg} = 0.7$. The collapse is quite good; however, in order to utilize the constitutive equation \eqref{segS_flux_T_seg} in practice, an additional constitutive equation that gives the granular temperature field in terms of other continuum quantities--such as strain-rate, stress, and fluidity--is needed. Recent work by \citet{kim2020} offers a path forward on this point. In summary, we acknowledge that the possibility for an alternative form for the segregation flux $w^{\rm seg}_i$ remains and that future work involving additional flow geometries will be required to conclusively judge which constitutive equation is the most predictive.

Another question regarding the constitutive equation for the segregation flux is the $c^{\rm l}$ dependence of the pre-factor. We simply use a symmetric dependence $c^{\rm l}(1-c^{\rm l})$, while other researchers \citep[e.g.,][]{van2015,tunuguntla2017} invoke expressions that depend on both $c^{\rm l}$ and $d^{\rm l}/d^{\rm s}$ in a more complex manner. 

Finally, in this paper, we have focused on two simple flow geometries that have two important features: (1) the continuum fields are one-dimensional, only varying along one spatial direction, and (2) the pressure field is spatially uniform. In order to apply the proposed continuum model in more complex flow geometries, such as heap flows or split-bottom flow, two important steps are necessary. First, this paper solely considered shear-strain-rate-driven size-segregation. Now that a predictive continuum model for this mechanism has been established, it remains to return to pressure-gradient-driven size-segregation in order to incorporate this mechanism by introducing an additional flux contribution to \eqref{flux_decomp_seg} to obtain a more general model. Second, a robust numerical implementation of the complex system of coupled equations that is capable of addressing problems in general geometries, involving multi-dimensional continuum fields, is needed. One possible approach is to utilize the finite-element method, as in previous work involving the NGF model \citep{henann2016}. These steps will be addressed in future works.

\section*{Acknowledgements}
This work was supported by funds from NSF-CBET-1552556.

\appendix
\section{Discrete-element method simulations and coarse-graining procedures}

\subsection{Simulated granular systems}\label{app_granularsystem}
We consider two types of simulated granular systems: two-dimensional systems consisting of a dense collection of circular disks and three-dimensional systems consisting of a dense collection of spheres. We consider bidisperse systems and denote the mean diameter of the large particles as $d^{\rm l}$ and the mean diameter of the small particles as $d^{\rm s}$ for both types. For both disks and spheres, we take $d^{\rm l}=3$\,mm and $d^{\rm s}=2$\,mm for the base case, so that $d_{\rm l}/d_{\rm s}=1.5$. For both large and small particles, the diameters of individual particles are chosen from a uniform distribution over the range of $\pm10\%$ of the respective mean diameters to prevent crystallization. In the two-dimensional granular system, $\rho_{\rm s}$ denotes the grain-material area-density, which we take to be $\rho_s = 3.26 \,{\rm kg/m^2}$ for both large and small disks, and in the three-dimensional granular system, $\rho_{\rm s}$ denotes the grain-material volume-density, which is taken to be $\rho_{\rm s} = 2450\, {\rm kg/m^3}$ for both large and small spheres to eliminate density-based segregation. For two-dimensional systems, the mass of the large disks is given by $ m^{\rm l}= ({\pi}/{4}) \rho_{\rm s} (d^{\rm l})^2$, and the mass of the small disks is given as $m^{\rm s}= ({\pi}/{4}) \rho_{\rm s} (d^{\rm s})^2$, so that the characteristic grain mass is $m=c^{\rm l}_0 m^{\rm l} + (1-c^{\rm l}_0)m^{\rm s}$, where $c^{\rm l}_0$ the initial concentration of the large grains. Similarly, for three-dimensional systems, the mass of the large spheres is $ m^{\rm l}= ({\pi}/{6}) \rho_{\rm s} (d^{\rm l})^3$, and the mass of the small spheres is $m^{\rm s}= ({\pi}/{6}) \rho_{\rm s} (d^{\rm s})^3$, so that the characteristic grain mass  is $m=c^{\rm l}_0 m^{\rm l} + (1-c^{\rm l}_0)m^{\rm s}$.

For the grain interaction model, the interaction force is given through a spring/dashpot contact law that accounts for elasticity, damping, and sliding friction \citep{dacruz2005,koval2009,kamrin2014,zhang2017}. The normal contact force $F_{\rm n}$ is given linearly through the normal component of the contact overlap, denoted by $\delta_{\rm n}$,  with  stiffness $k_{\rm n}$ and the relative normal velocity, denoted by $\dot{\delta}_{\rm n}$, with damping coefficient $g_{\rm n}$ as $F_{\rm n}=k_{\rm n}\delta_{\rm n} + g_{\rm n} \dot{\delta}_{\rm n}$ whenever $\delta_{\rm n}\ge0$ and $F_{\rm n}=0$ whenever $\delta_{\rm n}<0$. The normal damping coefficient is  given by $g_{\rm n} = \sqrt{m k_{\rm n}}(-2 \ln e)/\sqrt{2(\pi^2 + \ln^2 e)}$ where $e$ is the coefficient of restitution for binary collisions and $m$ is the characteristic grain mass discussed above. We denote the tangential stiffness and damping coefficient as $k_{\rm t}$ and $g_{\rm t}$, respectively, and take $g_{\rm t}=0$, so that the tangential force is given as  $F_{\rm t} = k_{\rm t} \delta_{\rm t}$ whenever $\delta_{\rm n}\ge0$, where $\delta_t$ is the tangential component of the contact displacement. The magnitude of the tangential component of the contact force is limited by Coulomb friction, which depends on the inter-particle friction coefficient $\mu_{\rm surf}$. Thus, the  parameter set $\{k_{\rm n}, k_{\rm t}, e, \mu_{\rm surf}\}$ fully describes the interaction properties. Throughout, the normal stiffness $k_{\rm n}$ is taken to be sufficiently large so that grains behave as stiff and nearly rigid. For two-dimensional systems, $k_{\rm n}/P > 10^4$, and for three-dimensional systems, $k_{\rm n}/P \bar{d}_0 > 10^4$, where $P$ is the characteristic confining pressure for a given configuration (force per unit length in two dimensions and force per unit area in three dimensions) and $\bar{d}_0=c^{\rm l}_0d^{\rm l} + (1-c^{\rm l}_0)d^{\rm s}$ is the characteristic grain size. In the stiff grain regime, the only interaction parameter that significantly affects the rheology of dense grains is $\mu_{\rm surf}$, which we have kept constant as $\mu_{\rm surf}=0.4$ throughout. The other parameters, namely the ratio $k_{\rm n}/k_{\rm t}$ and the restitution coefficient $e$, have negligible effects on the rheology \citep{kamrin2014} and thus the segregation dynamics in the stiff particle regime, so we maintain $k_{\rm t}/k_{\rm n} = 1/2$ and $e=0.1$ throughout. Finally, the open-source software LAMMPS \citep{lammps} is used to numerically integrate the equations of motion for each particle, and we restrict the time step for numerical integration to be $0.01$-$0.1$ of the binary collision time $\tau_{\rm c} = \sqrt{m\left(\pi^2 + \ln^2 e\right)/4 k_{\rm n}}$ for stability and accuracy of the simulation results. 

\subsection{Averaging methods}\label{app_average}

In this appendix, we describe the spatial and temporal averaging methods utilized to extract continuum fields from our DEM data. We begin with the spatial averaging procedure for a given snapshot of DEM data at a time $t$. All flows considered in this work are one-dimensional, in which the continuum fields vary only along one direction--i.e., along the $z$-direction in simple shear flow (Fig.~\ref{fig2_SegS}), along the $x$-direction in vertical chute flow (Fig.~\ref{fig4_SegS}), and along the $r$-direction in annular shear flow (Fig.~\ref{fig6_SegS})--and periodic along all other directions. Here, we describe the procedure for vertical chute flow of disks in detail, which may be straightforwardly adapted to the other flow geometries. We utilize a bin-based coarse-graining process, in which we construct a slender rectangle that spans the simulation domain along the $z$-direction and is centered at a given $x$-position with a finite width along the $x$-direction.\footnote{For annular shear flow of disks, the bins are thin, annular rings that are centered at a given $r$-position with a finite thickness along the $r$-direction.} Then, we assign each intersected grain $i$ a weight $A_i$, defined as the area of the grain $i$ inside the bin. Following \citet{tunuguntla2017} for a bidisperse system, we denote the sets of large and small grains intersected by the bin as ${\cal F}^{\rm l}$ and ${\cal F}^{\rm s}$, respectively, so that the set of all grains intersected by the bin is ${\cal F} = {\cal F}^{\rm l} \cup {\cal F}^{\rm s}$ with ${\cal F}^{\rm l} \cap {\cal F}^{\rm s} = \varnothing$.  The instantaneous solid area fraction field for species $\alpha$ is $\phi^{\alpha}(x,t) = (\sum_{i \in {\cal F}^\alpha} A_i)/A$, where $A$ is the total area of the bin, and the corresponding concentration field for species $\alpha$ is $c^{\alpha}(x,t) = \phi^{\alpha}(x,t)/\phi(x,t)$ with $\phi(x,t)=\phi^{\rm l}(x,t)+\phi^{\rm s}(x,t)$. With the instantaneous velocity of each grain $i$ denoted as ${\bf v}_i(t)$, the instantaneous velocity field is ${\bf v}(x,t)=(\sum_{i \in {\cal F}} A_i{\bf v}_i(t))/(\sum_{i \in {\cal F}} A_i)$.\footnote{We note that the definition of the instantaneous velocity field is consistent with first defining species-specific velocity fields--i.e., ${\bf v}^\alpha(x,t) = (\sum_{i\in{\cal F}^\alpha} A_i {\bf v}_i(t))/(\sum_{i \in {\cal F}^\alpha} A_i)$--and then calculating the mixture-level field--i.e., ${\bf v}(x,t) = c^{\rm l}(x,t) {\bf v}^{\rm l}(x,t) + (1 - c^{\rm l}(x,t)){\bf v}^{\rm s}(x,t)$.}  Likewise, with the instantaneous stress tensor associated with grain $i$ defined as $\boldsymbol{\sigma}_i(t) = (\sum_{j\ne i} {\bf r}_{ij}\otimes {\bf f}_{ij})/(\pi d_i^2/4)$, where ${\bf r}_{ij}$ is the position vector from the center of grain $i$ to the center of grain $j$, ${\bf f}_{ij}$ is the contact
force applied on grain $i$ by grain $j$, and $d_i$ is the diameter of grain $i$, the instantaneous stress field is $\boldsymbol{\sigma}(x,t) = (\sum_{i \in {\cal F}} A_i\boldsymbol{\sigma}_i(t))/A$. 

In vertical chute flow of spheres, we utilize a similar spatial coarse-graining approach. Instead of two-dimensional, rectangular bins, we use three-dimensional, rectangular-cuboidal bins that span the simulation domain along the $y$- and $z$-directions and are centered at a given $x$-position with a finite width along the $x$-direction. Then, the weight for each grain $i$ is the volume $V_i$ of the grain $i$ inside the bin. The instantaneous solid volume fraction field for species $\alpha$ is $\phi^\alpha(x,t) = (\sum_{i \in {\cal F}^\alpha} V_i)/V$, where $V$ is the total volume of the bin; the instantaneous concentration field for species $\alpha$ is $c^{\alpha}(x,t) = \phi^{\alpha}(x,t)/\phi(x,t)$; the instantaneous velocity field is ${\bf v}(x,t)=(\sum_{i \in {\cal F}} V_i{\bf v}_i(t))/(\sum_{i \in {\cal F}} V_i)$; the instantaneous stress tensor associated with grain $i$ is $\boldsymbol{\sigma}_i(t) = (\sum_{j\ne i} {\bf r}_{ij}\otimes {\bf f}_{ij})/(\pi d_i^3/6)$; and the instantaneous stress field is $\boldsymbol{\sigma}(x,t) = (\sum_{i \in {\cal F}} V_i\boldsymbol{\sigma}_i(t))/V$. 

Our analysis of the size-segregation process in this paper depends on obtaining accurate and high-resolution coarse-grained $c^{\rm l}$ fields from the DEM data. Therefore, the choices of the bin width and the spatial resolution of the bins are crucial. Throughout, we take a bin width of $4\bar{d}_0$ and a spatial resolution of roughly $0.1\bar{d}_0$ for both disks and spheres. Note that for these choices, adjacent bins overlap. We have ensured that a bin width of $4\bar{d}_0$ is sufficiently small so that the coarse-grained data is not over-smoothed. Specifically, we have tested that the coarse-grained velocity and stress fields are insensitive to the choice of bin width over the range of $0$ to $9.6\bar{d}_0$. In the limit that the bin width goes to zero, the coarse-graining procedure reduces to that utilized successfully in the literature for the velocity and stress fields for both disks \citep[e.g.,][]{dacruz2005,koval2009} and spheres \citep[e.g.,][]{zhang2017,kim2020}. However, as shown by \citet{weinhart2013}, applying a coarse-graining procedure with a small or zero bin width to the $\phi$ field--and hence the $\phi^{\rm l}$, $\phi^{\rm s}$, $c^{\rm l}$, and $c^{\rm s}$ fields--will lead to spatial fluctuations due to particle layering near the walls. We have ensured that a bin width of $4\bar{d}_0$ is sufficiently large so that these layering effects are not observed in the $c^{\rm l}$ field--i.e., we are within the ``plateau range'' \citep{weinhart2013} of bin widths that produce bin-width-independent, coarse-grained continuum fields. We note that when plotting spatiotemporal contours of the concentration fields, profiles of the concentration fields, and profiles of the velocity fields (e.g., Fig.~\ref{fig4_SegS}), we truncate the coarse-grained DEM data from bins centered within one-half of a bin-width ($2\bar{d}_0$) from the walls. Furthermore, to ensure that the DEM data collapses used to determine the dimensionless material parameter $C_{\rm seg}^{\rm S}$  (i.e., Figs.~\ref{fig5_SegS}, \ref{fig7_SegS}, and \ref{fig10_SegS}) are representative of bulk behavior and not wall effects, we use a more conservative criterion and do not include DEM data from bins within $6\bar{d}_0$ of the walls. 

The collapses of Figs.~\ref{fig5_SegS}, \ref{fig7_SegS}, and \ref{fig10_SegS} are obtained in the long-time regime, in which the fields evolve slowly in time. Since the flow is quasi-steady, the instantaneous concentration and velocity fields are simply arithmetically averaged in time (using 152 instantaneous snapshots for vertical chute flow of disks, 1000 snapshots for vertical chute flow of spheres, and 144 snapshots for annular shear flow of disks) to obtain fields that only depend on the spatial coordinate. Then, the necessary first and second-order spatial derivatives of the field quantities (e.g., $\partial c^{\rm l}/\partial x$, $\dot\gamma = \partial v_z/\partial x$, and $\partial \dot\gamma/\partial x = \partial^2 v_z/\partial x^2$ for vertical chute flow) are obtained from these time-averaged fields. We apply a spatial derivative filter to the time-averaged DEM fields in order to obtain accurate estimates of the spatial derivatives. We have tested using both cutoff Gaussian functions and Lucy functions \citep{weinhart2013,tunuguntla2016} for the kernel function of the derivative filter as well as a range of kernel function widths to ensure that the reported results in this study are independent of these choices. 

Unlike for the steady concentration and velocity fields, which allow for arithmetic time-averaging, the transient concentration fields for dense flows of disks (e.g., Fig.~\ref{fig4_SegS}(d)) are time-averaged in a slightly different manner using a cutoff Gaussian filter. In particular, this process is performed by applying a normalized, cutoff Gaussian time filter to the DEM data at each $x$-position. Denoting the standard deviation of the Gaussian kernel function as $\sigma_{\rm t}$, so that the cutoff time-width of the Gaussian kernel is $6\sigma_{\rm t}$, the time-smoothed field quantity at a given $x$-position and time $t$ is then given by the convolution of the DEM data over a time-period of $6\sigma_{\rm t}$, centered at time $t$, with the cutoff Gaussian kernel. We have tested a range of kernel widths $\sigma_{\rm t}$ to ensure that the coarse-grained concentration fields appearing in this paper are insensitive to this choice. For the transient concentration fields for dense flows of spheres (second column of Fig.~\ref{fig_SegS_spheres}), no additional time-smoothing is needed, and the concentration fields are simply instantaneous snapshots in time. This is possible primarily because spatial smoothing is being done over a greater volume and a larger number of grains, and therefore the instantaneous concentration fields are relatively more smooth. 

\section{Diffusion flux consistency test}
\label{app_diff_seg}
Our process for determining the segregation flux--and hence the material parameter $C^{\rm S}_{\rm seg}$--is based on the assumption that the segregation and diffusion fluxes balance in the quasi-steady regime. Therefore, it is essential that the dimensionless material parameter $C_{\rm diff}$ appearing in the constitutive equation for the diffusion flux \eqref{diff_flux_seg} has been accurately determined, so that the coarse-grained diffusion flux is accurate. In Section~\ref{sec_diff_seg}, we determined $C_{\rm diff}$ for dense flows of frictional disks to be 0.20 using mean square displacement data from DEM simulations of simple shear flow of a well-mixed bidisperse granular system. In this Appendix, we perform an independent consistency check that tests whether the constitutive equation for the diffusion flux \eqref{diff_flux_seg} using this fitted value of $C_{\rm diff}$ for disks is capable of predicting the evolution of the $c^{\rm l}$ field in a diffusion-dominated problem.

\begin{figure}[!t]
\begin{center}
\includegraphics{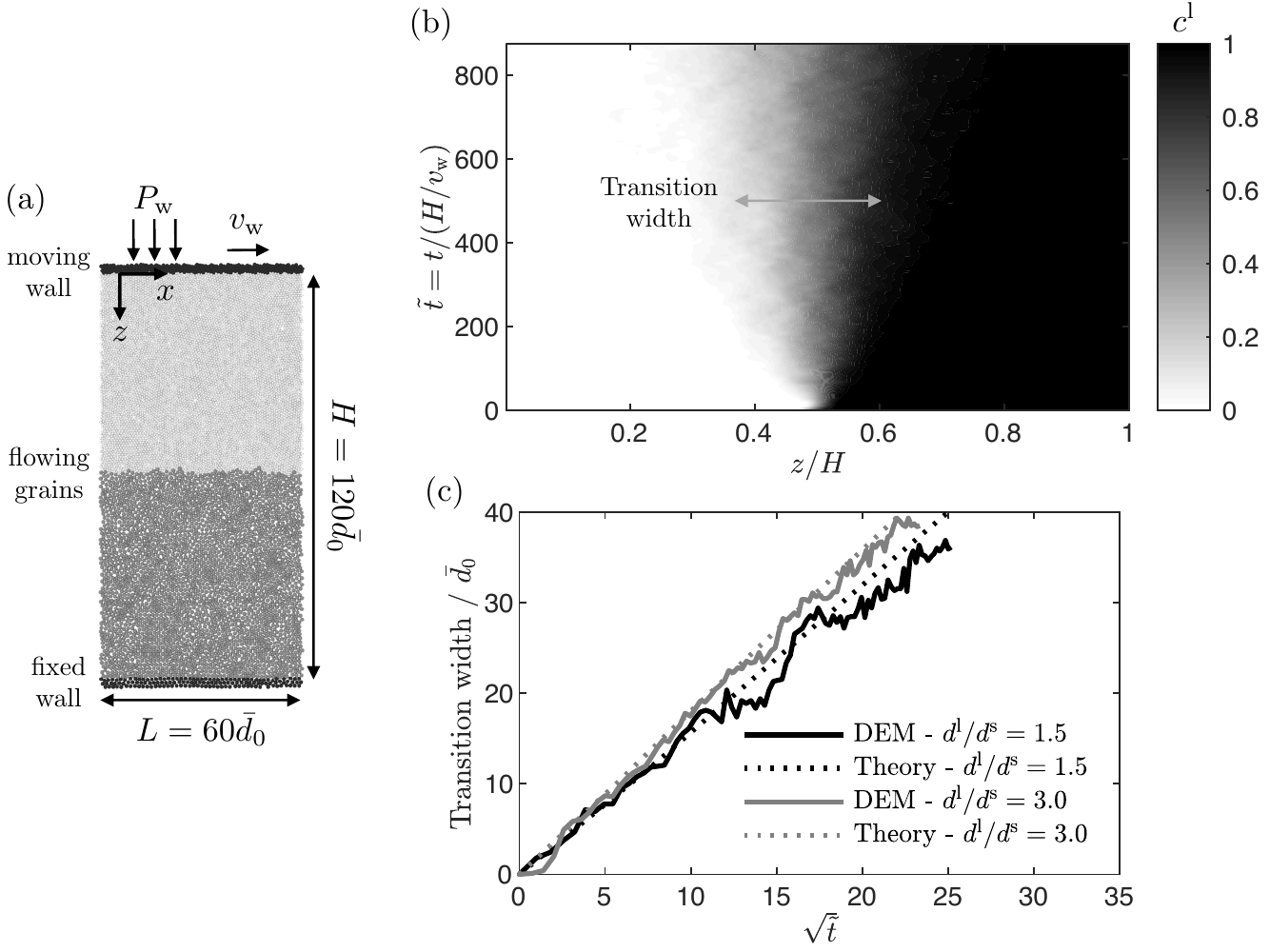}
\end{center}
\caption{(a) Initially-segregated configuration for two-dimensional DEM simulation of bidisperse simple shear flow with $d^{\rm l}/d^{\rm s}=1.5$ and $8649$ flowing grains. Upper and lower layers of black grains denote rough walls. Dark gray grains indicate large flowing grains, and light gray grains indicate small flowing grains. A 10\% polydispersity is utilized for each species to prevent crystallization. (b) Spatiotemporal evolution of the large grain
concentration field, illustrating the transition width that grows with time. (c) Normalized transition width versus square root of normalized time $\tilde{t} = t/(H/v_{\rm w})$.}\label{fig1_appen_seg}
\end{figure}

Consider homogeneous simple shear flow of an initially-segregated system with large grains (dark gray) on the bottom and small grains (light gray) on the top, as shown in Fig.~\ref{fig1_appen_seg}(a) for the case of $d^{\rm l}/d^{\rm s}=1.5$. The rectangular domain has a length of $L=60\bar{d}_0$ in the $x$-direction and a height of $H=120\bar{d}_0$ in the $z$-direction. As in Sections~\ref{subsec_rheo_seg} and \ref{sec_diff_seg}, shearing along the $x$-direction and normal stress along the $z$-direction are applied by the walls. We perform DEM simulations of simple shearing for a nominal inertial number of $(v_{\rm w}/H)\sqrt{\bar{d}_0^2\rho_{\rm s}/P_{\rm w}} = 0.1$. We run the DEM simulation starting from the initially-segregated configuration, and the spatiotemporal evolution of the coarse-grained $c^{\rm l}$-field is shown in Fig.~\ref{fig1_appen_seg}(b) for $d^{\rm l}/d^{\rm s}=1.5$. We observe that the interface between large and small grains, which is initially sharp, becomes diffuse with a transition width that grows with time. We define a transition width at a given point in time as the distance between the positions at which $c^{\rm l}$ equals 0.1 and 0.9 in snapshots of the spatial $c^{\rm l}$ profile. This transition width as a function of the square-root of the dimensionless time $\tilde{t} = t/(H/v_{\rm w})$ is plotted in Fig.~\ref{fig1_appen_seg}(c) as solid curves for grain-size-ratios of $d^{\rm l}/d^{\rm s}=1.5$ and 3.0, displaying roughly linear behavior--typical of diffusive behavior--with a slight dependence on $d^{\rm l}/d^{\rm s}$. 

Next, we apply the continuum model for the evolution of $c^{\rm l}$ \eqref{segS_model_seg} to this problem. As for planar shear flow of a well-mixed bidisperse system (Fig.~\ref{fig2_SegS}), no pressure gradient is present. Therefore, the evolution of $c^{\rm l}$ is governed by \eqref{segS_model_seg}:  
\begin{equation}\label{rev_mix_model_seg}
\dfrac{d{c}^{\rm l}}{dt} + \dfrac{\partial}{\partial z} \left( -C_{\rm diff} \bar{d}^2 \dot{\gamma} \dfrac{\partial c^{\rm l}}{\partial z} + C^{\rm S}_{\rm seg} \bar{d}^2 c^{\rm l}(1-c^{\rm l}) \dfrac{\partial \dot{\gamma}}{\partial z} \right) = 0 ,
\end{equation}
where $\bar{d} = c^{\rm l}d^{\rm l} + (1 - c^{\rm l})d^{\rm s}$. In this flow configuration, the shear-strain-rate is approximately constant. Therefore, the shear-strain-rate-gradient is approximately zero throughout, and the diffusion flux is the dominant flux, which acts to remix the flowing grains. This may be understood in the context of the local inertial rheology. Since the stress ratio $\mu$ is spatially-constant in homogeneous planar shear, the resulting inertial number field $I$ is also spatially constant. Since the inertial number depends on both $\bar{d}$ and $\dot\gamma$, spatial variation in $\bar{d}$ leads to spatial variation in $\dot\gamma$. This spatial variation in $\dot\gamma$ is slight, and consequently, the magnitude of the diffusion flux is at each point in space is much greater than the magnitude of the segregation flux. While the effect of segregation is small, we still include the shear-strain-rate-gradient-driven segregation flux with $C_{\rm seg}^{\rm S}= 0.23$ and solve \eqref{rev_mix_model_seg} when analyzing the problem shown in Fig.~\ref{fig1_appen_seg}(a). We note that due to the small effect of segregation and the dependence of $\bar{d}$ on $c^{\rm l}$, \eqref{rev_mix_model_seg} is similar to but not exactly the same as the linear diffusion equation in one dimension, so the solution is close to but not exactly an error function.

We obtain predictions for the evolution of the $c^{\rm l}$-field by solving \eqref{rev_mix_model_seg} using the fully-segregated initial condition for $c^{\rm l}(z,t=0)$, no-flux boundary conditions at $z=0$ and $z=H$, a spatially-constant value of inertial number $I$ consistent with that prescribed in the DEM simulations, a given grain-size-ratio $d^{\rm l}/d^{\rm s}$, and $C_{\rm diff}=0.20$. We note that since $I$ is taken to be spatially-constant, $\dot\gamma = (I/\bar{d})\sqrt{P_{\rm w}/\rho_{\rm s}}$ varies slightly in space due to the $c^{\rm l}$-dependence of $\bar{d}$. We extract the transition width as a function of time from continuum simulation results for $d^{\rm l}/d^{\rm s}=1.5$ and 3.0 and include these results in Fig.~\ref{fig1_appen_seg}(c) as dashed lines. The continuum model predictions agree well with the DEM data and are even capable of capturing the small difference due to the grain-size-ratio. This result indicates that the expression for the diffusion flux \eqref{diff_flux_seg} and the fitted material parameter value $C_{\rm diff}=0.20$ are indeed consistent with DEM data for disks.

\end{document}